\documentclass[aps,pre,showpacs,superscriptaddress,groupedaddress]{revtex4} 
\usepackage{graphicx}  
\usepackage{dcolumn}   
\usepackage{bm}        
\usepackage{amssymb}   

\begin{document}

\title{Selection of sequence motifs and generative Hopfield-Potts
  models for protein families} 
\author{Kai Shimagaki}
\affiliation{Sorbonne Universit\'{e}, CNRS, Institut de Biologie
  Paris-Seine, Laboratoire de Biologie Computationnelle et
  Quantitative -- LCQB Paris, France} 
\author{Martin Weigt}
\affiliation{Sorbonne Universit\'{e}, CNRS, Institut de Biologie
  Paris-Seine, Laboratoire de Biologie Computationnelle et
  Quantitative -- LCQB Paris, France} 

\date{\today}


\begin{abstract}
Statistical models for families of evolutionary related proteins have
recently gained interest: in particular pairwise Potts models, as
those inferred by the Direct-Coupling Analysis, have been able to
extract information about the three-dimensional structure of folded
proteins, and about the effect of amino-acid substitutions in
proteins. These models are typically requested to reproduce the one-
and two-point statistics of the amino-acid usage in a protein
family, {\em i.e.}~to capture the so-called residue conservation and
covariation statistics of proteins of common evolutionary
origin. Pairwise Potts models are the maximum-entropy models achieving
this. While being successful, these models depend on huge numbers of
{\em ad hoc} introduced parameters, which have to be estimated from
finite amount of data and whose biophysical interpretation remains
unclear. Here we propose an approach to parameter reduction, which is
based on selecting collective sequence motifs. It naturally leads to
the formulation of statistical sequence models in terms of 
Hopfield-Potts models. These models can be accurately inferred using a
mapping to restricted Boltzmann machines and persistent contrastive
divergence. We show that, when applied to protein data, even 20-40
patterns are sufficient to obtain statistically close-to-generative
models. The Hopfield patterns form interpretable sequence motifs and
may be used to clusterize amino-acid sequences into functional 
sub-families. However, the distributed collective nature of these motifs
intrinsically limits the ability of Hopfield-Potts models in predicting
contact maps, showing the necessity of developing models going beyond 
the Hopfield-Potts models discussed here. 
\end{abstract}

\maketitle

\section{Introduction}
\label{sec:intro}

Thanks to important technological advances, exemplified in particular
by next-generation sequencing, biology is currently undergoing a deep
transformation towards a data-rich science. As an example, the number
of available protein sequences deposited in the Uniprot database was
about 1 million in 2004, crossed 10 millions in 2010, and 100 millions
in 2018, despite an important reorganization of the database in 2015
to reduce redundancies and thus limit database size
\cite{uniprot2018uniprot}. On the contrary, proteins with detailled
experimental knowledge are contained in the manually annotated
SwissProt sub-database of Uniprot. While their number remained almost
constant and close to 500,000 over the last decade, the knowledge
about these selected proteins has been continuously extended and
updated. 

This fastly growing wealth of data is presenting both a challenge and
an opportunity for data-driven modeling approaches. It is a challenge,
because for less than 0.5\% of all known protein sequences at least
some knowledge going beyond sequence is available. Applicability of
standard supervised machine-learning approaches is thus frequently
limited. However, more importantly, it is an opportunity since
protein-sequence databases like Uniprot are not large sets of
unrelated random sequences, but contain structured, functional
proteins resulting from natural evolution. 

In particular, protein sequences can be classified into so-called {\em
  homologous protein families} \cite{finn2016pfam}. Each family
contains protein 
sequences, which are believed to share common ancestry in
evolution. Such homologous sequences typically show very similar
three-dimensional folded structures, and closely related biological
functions. Put simply, they can be seen as equivalent proteins in
different species, or in different pathways of the same
species. Despite this high level of structural and functional
conservation, homologous proteins may differ in more than 70-80\%
of their amino-acids. Detecting homology between a currently
uncharacterized protein and a well-studied one
\cite{eddy2009new,remmert2012hhblits}  is therefore the most
important means for computational sequence annotation, including
protein-structure prediction by homology modeling 
\cite{schwede2003swiss,webb2014comparative}.   

To go beyond such knowledge transfer, we can explore the observable
sequence variability between homologous proteins, since it contains on
its own important information about the evolutionary constraints
acting on proteins to conserve their structure and function
\cite{de2013emerging}. Typically very few random mutations do actually
destabilize proteins or interrupt their function. Some positions need
to be highly conserved, while others are permissive for multiple
mutations. Observing sequence variability across entire homologous
protein families, and relating them to protein structure, function,
and evolution, is therefore an important task \cite{cocco2018inverse}. 

Over the last years, {\em inverse statistical physics}
\cite{nguyen2017inverse} has played an 
increasing role in solving this task. Methods like 
Direct-Coupling Analysis (DCA)
\cite{weigt2009identification,morcos2011direct} or related approaches
\cite{balakrishnan2011learning,jones2012psicov} allow for 
predicting protein structure
\cite{marks2011protein,ovchinnikov2017protein}, mutational effects
\cite{chakraborty2014hiv,morcos2014coevolutionary,figliuzzi2015coevolutionary}
and protein-protein 
interactions \cite{szurmant2018inter}. However, many of these methods
depend on huge numbers  of typically {\em ad hoc} introduced
parameters, making these methods data-hungry and susceptible to
overfitting effects.

In this paper, we describe an attempt to substantially reduce the
amount of parameters, and to select them systematically using sequence
data. Despite this parameter reduction, we aim at so-called {\it
  generative} statistical models: samples drawn from these models
should be statistically similar to the real data, even if similarity
is evaluated using statistical measures, which were not used to infer
the model  from data.

To this aim, we first review in Sec.~\ref{sec:summary} some important
points about protein-sequence data, maximum-entropy models of these
data in general, and profile and DCA models in particular. In
Sec.~\ref{sec:cvs_hp}, we introduce a way for rational selection of
so-called sequence motifs, which generalizes maximum-entropy
modeling. The resulting Hopfield-Potts models are mapped to Restricted
Boltzmann Machines (recently introduced independently for proteins in
\cite{tubiana2019learning}) in Sec.~\ref{sec:inference} to enable
efficient model inference and interpretation of the model
parameters. Sec.~\ref{sec:proteins} is dedicated to the application of
this scheme to some exemplary protein families. The Conclusion and
Outlook in Sec.~\ref{sec:conclusion} is followed by some technical
appendices. 

\section{A short summary: Sequence families, MaxEnt models and DCA} 
\label{sec:summary}

To put our work into the right context, we need to review shortly some
published results about the statistical models of protein
families. After introducing the data format, we summarize the 
maximum-entropy approach typically used to justify the use of
Boltzmann distributions for protein families, together with some
important shortcomings of this approach. Next we give a concise
overview over two different types of maximum-entropy models -- profile
models and direct-coupling analysis --  which are currently used for
protein sequences. For all cases we discuss the strengths and
limitations, which have motivated our current work.

\subsection{Sequence data}

Before discussing modeling strategies, we need to properly define what
type of data is used. Sequences of homologous proteins are used in the
form of {\em multiple-sequence alignments} (MSA), {\em
  i.e.}~rectangular matrices $(A_i^m)_{i=1,...,L}^{m=1,...,M}$. Each of the 
rows $m=1,...,M$ of this matrix contains one aligned protein sequence
$\underline A^m = (A_1^m,...,A_L^m)$ of length $L$. In the context of 
MSA, $L$ is also called the alignment width, $M$ its depth. Entries in
the matrix come from the alphabet ${\cal A} = \{-,A,C,...,Y\}$
containing the 20 natural amino acids and the alignment gap
``--''. Throughout this paper, the size of the alphabet will be
denoted by $q=21$. In practice we will use a numerical version of the
alphabet, denoted by $\{1,...,q\}$, but we have to keep in mind that
variables are categorical variables, {\em i.e.} there is no linear
order associated to these numerical values.

The Pfam database \cite{finn2016pfam}  currently (release 32.0) lists
almost 18,000 protein families. Statistical modeling is most
successful for large families, which contain between $10^3$ and $10^6$
sequences. Typical lengths span the range $L=30-500$.

\subsection{Maximum-entropy modeling}

The aim of statistical modeling is to represent each protein family by
a function $P(\underline A)$, which assigns a probability to each
sequence $\underline A \in {\cal A}^L$, {\it i.e.}~to each sequence formed 
by $L$ letters from the amino-acid alphabet ${\cal A}$. Obviously the
number of sequences even in the largest MSA is much smaller than the
number $q^L-1$ of a priori independent parameters characterizing
$P$. So we have to use clever parameterizations for these models.

A commonly used strategy is based on the maximum-entropy (MaxEnt)
approach \cite{jaynes1957information}. It start from any number $p$ of
observables, 
\begin{equation}
\label{eq:observable}
{\cal O}^\mu :  {\cal A}^L \to \mathbb{R}\ ,\ \ \ \  \mu=1,...,p,
\end{equation}   
which assign real numbers to each sequence. Only the values of these
observables for the sequences in the MSA  $(\underline A^m)$ go into
the MaxEnt models. More precisely, we require the model to reproduce
the empirical mean of each observable over the data:
\begin{equation}
\label{eq:consistency}
\forall \mu=1,...p\ : \ \ \ \ \sum_{\underline A \in {\cal A}^L}
P(\underline A)\, {\cal  O}^\mu(\underline A) 
= \frac 1M \sum_{m=1}^M {\cal  O}^\mu(\underline A^m)\ .
\end{equation}
In a more compact notation, we write $\langle {\cal  O}^\mu \rangle_P
= \langle {\cal  O}^\mu \rangle_{\rm MSA}$. Besides this consistency
with the data, the model should be as unconstrained as possible. Its
entropy has therefore to be maximized,
\begin{equation}
\label{eq:maxent}
-\sum_{\underline A \in {\cal A}^L} P(\underline A) \log P(\underline A)
\longrightarrow \max\ .
\end{equation}
Imposing the constraints in Eq.~(\ref{eq:consistency}) via Lagrange
multipliers $\lambda_\mu, \mu=1,...,p,$ we immediately find that
$P(\underline A)$ assumes a Boltzmann-like exponential form
\begin{equation}
\label{eq:maxent_P}
P(\underline A) = \frac 1Z \exp\left\{
\sum_{\mu=1}^p \lambda_\mu {\cal  O}^\mu(\underline A)
\right\} \ .
\end{equation}
Model inference consists in fitting the Lagrange multipliers such that
Eqs.~(\ref{eq:consistency}) are satisfied. The partition function $Z$
guarantees normalization of $P$.

MaxEnt relates observables and the analytical form of the probability
distribution, but it does not provide any rule how to select
observables. Frequently prior knowledge is used to decide, which
observables are ``important'' and which not. More systematic
approaches therefore have to address at least the following two
questions:
\begin{itemize}
\item Are the selected observables {\em sufficient}? In the best case,
  model $P$ becomes {\em generative}, {\em i.e.} sequences $\underline A$
  sampled from $P$ are statistically indistinguishable from the
  natural sequences in the MSA $(\underline A^m)$ used for model
  learning. While this is hard to test in full generality, we can select
  observables {\em not} used in the construction of the model, and
  check if their averages in the model and over the input data
  coincide. 
\item Are the selected observables {\em necessary}? Would it be
  possible to construct a parameter-reduced, thus more parsimonious
  model of same quality? This question is very important due to at
  least two reasons: (a) the most parsimonious model would allow for
  identifying a minimal set of evolutionary constraints acting on
  proteins, and thus offer deep insight into protein evolution; and
  (b) a reduced number of parameters would allow to reduce overfitting
  effects, which result from the limited availability of data ($M\ll q^L$).
\end{itemize}
While there has been promising progress in the first question,
cf.~the next two subsections, our
work attempts to approach both questions simultaneously, thereby going
beyond standard MaxEnt modeling.

To facilitate the further discussion, two important technical points have
to be mentioned. First, MaxEnt leads to a family of so-called
exponential models, where the exponent in Eq.~(\ref{eq:maxent_P}) is
{\em linear} in the Lagrange multipliers $\lambda_\mu$, which parameterize
the family. Second, MaxEnt is intimately related to maximum
likelihood. When we postulate Eq.~(\ref{eq:maxent_P}) for the
mathematical form of model $P(\underline A)$, and when we maximize the
log-likelihood
\begin{equation}
{\cal L}( \{\lambda_\mu\}\, |\, (A_i^m) )
= \sum_{m=1}^M \log P(\underline A^m )
\end{equation}
with respect to the parameters $\lambda_\mu, \mu=1,...,P,$ we
rediscover Eqs.~(\ref{eq:consistency}) as the stationarity
condition. The particular form of $P(\underline A)$ guarantees that
the likelihood is convex, having only a unique maximum.

\subsection{Profile models}

The most successful approach in statistical modeling of biological
sequences are probably {\it profile models}
\cite{durbin1998biological}, which consider each MSA 
column ({\em i.e.} each position in the sequence) independently. The
corresponding observables are simply ${\cal O}^{ia}(\underline A) =
\delta_{A_i,a}$ for all positions $i=1,...,L$ and all amino-acid
letters $a\in {\cal A}$, with $\delta$ being the standard Kronecker
symbol. These observables thus just ask if in a sequence $\underline
A$, amino acid $a$ is present in position $i$. Their statistics
in the MSA is thus characterized by the fraction
\begin{equation}
\label{eq:fi}
f_i(a) = \frac 1M \sum_{m=1}^M \delta_{A_i^m,a}
\end{equation}
of sequences having amino acid $a$ in position $i$. Consistency of
model and data requires marginal single-site distributions of $P$ to
coincide with the $f_i$,
\begin{equation}
\label{eq:Pi}
\forall i=1,...,L, \forall A_i\in{\cal A}\ :\ \ \ \ \sum_{ \{A_j |
  j\neq i\}} P(\underline A) = f_i(A_i)\ .
\end{equation}
The MaxEnt model results as $P(\underline A)=\prod_{i=1}^L f_i(A_i)$, which
can be written as a factorized Boltzmann distribution 
\begin{equation}
\label{eq:profile}
P(\underline A) = \frac 1Z \exp\left\{
\sum_i h_i(A_i)
\right\}\ ,
\end{equation}
where the local fields equal $h_i(a)=\log f_i(a)$. Pseudocounts or
regularization can be used to avoid infinite negative parameters for
amino acids, which are not observed in some MSA column.

Profile models reproduce the so-called {\em conservation} statistics
of an MSA, {\em i.e.} the heterogenous usage of amino acids in the
different positions of the sequence. Conservation of a single or few
amino-acids in a column of the MSA is typically an indication of an
important functional or structural role of that position. Profile
models, frequently in their generalization to profile Hidden Markov
Models \cite{eddy1998profile,eddy2009new,remmert2012hhblits}, are used
for detecting homology of new sequences to protein families, for
aligning multiple sequences, and -- using the conserved structural
and functional characteristics of protein families -- indirectly for the
computational annotation of experimentally uncharacterized amino-acid 
sequences. They are in fact at the methodological basis of the
generation of the MSA used here.

Despite their importance in biological sequence analysis, profile
models are not generative. Biological sequences show significant
correlation in the usage of amino acids in different positions, which
are said to {\em coevolve} \cite{de2013emerging}. Due to their
factorized nature,  profile models are not able to reproduce these
correlations, and larger sets of observables have to be used to obtain
potentially generative sequence models. 

\subsection{Direct-Coupling Analysis}

The Direct-Coupling Analysis (DCA)
\cite{weigt2009identification,morcos2011direct} 
therefore includes also pairwise
correlations into the modeling. The statistical model
$P(\underline A)$ is not only required to reproduce the amino-acid
usage of single MSA columns, but also the fraction $f_{ij}(a,b)$ of
sequences having simultaneously amino acid $a$ in position $i$, and
amino acid $b$ in position $j$, for all $a,b\in {\cal A}$ and all
$1\leq i<j\leq L$:
\begin{eqnarray}
\label{eq:fij}
f_{ij}(a,b) &=& \frac 1M \sum_{m=1}^M \delta_{A_i^m,a}\, \delta_{A_j^m,b}
                \nonumber\\
&=& \sum_{\underline A\in {\cal A}^L} P(\underline A)\,
    \delta_{A_i,a}\, \delta_{A_j,b}\ .
\end{eqnarray}
The corresponding observables $\delta_{A_i,a}\,\delta_{A_j,b}$ are
thus products of pairs of observables used in profile models.

According to the general MaxEnt scheme described before, DCA leads to
a generalized $q$-states Potts model
\begin{equation}
\label{eq:potts}
P(\underline A) = \frac 1Z \exp\left\{
\sum_{i<j} J_{ij}(A_i,A_j) + \sum_i h_i(A_i)
\right\}
\end{equation}
with heterogeneous pairwise couplings $J_{ij}(a,b)$ and local fields
$h_i(a)$. The inference of parameters becomes computationally hard,
since the computation of the marginal distributions in
Eq.~(\ref{eq:fij}) requires to sum over $O(q^L)$ sequences. Many
approximation schemes have been proposed, including message-passing
\cite{weigt2009identification}, 
mean-field \cite{morcos2011direct}, Gaussian
\cite{jones2012psicov,baldassi2014fast} and pseudo-likelihood
maximization \cite{balakrishnan2011learning,ekeberg2013improved}
approximations. DCA and related global inference techniques have found
widespread applications in the prediction of protein structures, of
protein-protein interactions and of mutational effects, demonstrating
that amino-acid covariation as captured by the $f_{ij}$ contains
biologically valuable information. 

While these approximate inference schemes do not lead to generative
models -- not even the $f_i$ and the $f_{ij}$ are accurately reproduced
-- recently very precise but time-extensive inference schemes based on
Boltzmann-machine learning have been proposed
\cite{sutto2015residue,haldane2016structural,barton2016ace,figliuzzi2018pairwise}. 
Astonishingly, these models do not only reproduce the fitted 
one- and two-column statistics of the input MSA: also non-fitted
characteristics like the three-point statistics $f_{ijk}(a,b,c)$ or
the clustered organization of sequences in sequence space are
reproduced. These observations strongly suggest that pairwise Potts
models as inferred via DCA are generative models, {\em i.e.}~that the
observables used in DCA -- amino-acid occurrence in single positions
and in position pairs -- are actually defining a (close to) sufficient
statistics. In a seminal experimental work
\cite{socolich2005evolutionary},  the importance of respecting
pairwise correlations in amino-acid usage in generating small
artificial but folding protein sequences was shown.

However, DCA uses an enormous amount of parameters. There are
independent couplings for each pair of positions and amino acids. In
case of a protein of limited length $L=200$, the total number of
parameters is close to $10^8$. Very few of these parameters are
interpretable in terms of, {\em e.g.}, contacts between positions in
the three-dimensional protein fold. We would therefore expect that not
all of these observables are really important to statistically model
protein sequences. On the contrary, given the limited size
($M=10^3-10^4$) of most input MSA, the large number of parameters
makes overfitting likely, and quite strong regularization is
needed. It would therefore be important to devise parameter-reduced 
models, as proposed in \cite{cocco2013principal}, but without giving
up on the generative character of the inferred statistical models.

\section{From sequence motifs to the Hopfield-Potts model}
\label{sec:cvs_hp}

Seen the importance of amino-acid conservation in proteins, and of
profile models in computational sequence analysis, we keep
Eqs.~(\ref{eq:Pi}), which link the single-site marginals of
$P(\underline A)$ directly to the amino-acid frequencies $f_i(a)$ in
single MSA columns. Further more, we assume that the important
observables for our protein-sequence ensemble can be represented as
so-called {\em sequence motifs}
\begin{equation}
\label{eq:motif}
{\cal O}^\mu(\underline A) = \sum_i \omega_i^\mu(A_i)\ ,\ \ \ \mu=1,...,p,
\end{equation}
which are linear additive combinations of single-site terms. In
sequence bioinformatics, such sequence motifs are widely used, also
under alternative names like {\em position-specific scoring / weight
  matrices}, cf.~\cite{stormo1982use,van2007finding}. Note that, in
difference to the observables introduced before for profile or DCA
models, motifs constitute {\it collective observables} potentially
depending on the entire amino-acid sequence.

Let us assume for a moment that these motifs, or more specifically the
corresponding $\omega$-matrices, are known. We will address their
selection later. For any model $P$ reproducing the sequence profile,
{\em i.e.} for any model fulfilling Eqs.~(\ref{eq:Pi}), also the
ensemble average of the ${\cal O}^\mu$ is given,
\begin{equation}
\label{eq:1st_O}
\sum_{\underline A} P(\underline A)\, {\cal O}^\mu (\underline A) =
\sum_{i,a} \omega_i^\mu(a) f_i(a)\ .
\end{equation}
The empirical mean of these observables therefore does not contain
any further information about the MSA statistics beyond the profile
itself. The key step is to consider also the variance, or the second
moment, 
\begin{equation}
\label{eq:2nd_O}
\frac 1M \sum_{m} \left[ {\cal O}^\mu (\underline A^m)
  \right]^2 =
\sum_{i,j,a,b} \omega_i^\mu(a) \omega_j^\mu(b) f_{ij}(a,b)
\end{equation}
as a distinct feature characterizing the sequence variability in the
MSA, which has to be reproduced by the statistical model $P(\underline
A)$. This second moment actually depends on combinations of the
$f_{ij}$, which were introduced in DCA to account for the correlated
amino-acid usage in pairs of positions.

The importance of fixing this second moment becomes clear in a very
simple example: consider only two positions $\{1,2\}$ and two possible
letters $\{A,B\}$, which are allowed in these two positions. Let us
assume further that these two letters are equiprobable in these two
positions, {\em i.e.} $f_{1}(A) = f_{1}(B) = f_{2}(A) = f_{2}(B) =
1/2$. Assume further a single motif to be given by $\omega_{1}(A) =
\omega_{2}(A) = 1/2,\  \omega_{1}(B) = \omega_{2}(B) =
-1/2$. In this case, the mean of ${\cal O}$ equals zero. We
further consider two cases:
\begin{itemize}
\item {\em Uncorrelated positions:} In this case, all words
  $AA,AB,BA,BB$ are equiprobable. The second moment of ${\cal O}$ thus
  equals 1/2.
\item {\em Correlated positions:} As a strongly correlated example, only
  the two words $AA$ and $BB$ are allowed. The second moment of
  ${\cal O}$ thus equals 1.
\end{itemize}
We conclude that an increased second moment (or variance) of these
additive observables with respect to the uncorrelated case corresponds
to the preference of combinations of letters or entire words; this is
also the reason why they are the called sequence motifs.

Including therefore these second moments as conditions
into the MaxEnt modeling, our statistical model takes the shape
\begin{equation}
\label{eq:maxent_hp} 
P\left(\underline A\, |\, 
\{\lambda_\mu, h_i(a), \omega_i^\mu(a)\}\right) = \frac 1Z \exp\left\{
\sum_{\mu=1}^p \lambda_\mu \sum_{i,j=1}^L \omega_i^\mu(A_i)\omega_j^\mu(A_j) +
\sum_{i=1}^L h_i(A_i)
\right\}
\end{equation}
with Lagrange multipliers $\lambda_\mu, \mu=1,...,p,$ imposing means
(\ref{eq:2nd_O}) to be reproduced by the model, and $h_i(a),
i=1,...,L, a\in {\cal A},$ to impose Eqs.~(\ref{eq:Pi}).

\subsection{The Hopfield-Potts model: from MaxEnt to
  sequence-motif selection}

As mentioned before, an important limitation of MaxEnt models is that
they assume certain observables to be reproduced, but they do not
offer any strategy, how these observables have to be selected. In the
case of Eq.~(\ref{eq:maxent_hp}), this accounts in particular to
optimizing the values of the Langrange parameters $\lambda_\mu$ to
match the ensemble averages over $P(\underline A)$ with the sample
averages Eq.~(\ref{eq:2nd_O}) over the input MSA. As mentioned before,
this corresponds also to {\it maximizing the log-likelihood} of these
parameters given MSA and the $\omega$-matrices describing the motifs, 
\begin{equation}
{\cal L}( \{\lambda_\mu, h_i(a)\}\, |\, (A_i^m), \{\omega_i^\mu(a)\})
= \sum_{m=1}^M \log P(\underline A^m\, |\, 
\{\lambda_\mu, h_i(a), \omega_i^\mu(a)\} )\ .
\end{equation}

The important, even if quite straight forward {\em step from MaxEnt
modeling to Motif Selection} is to optimize the likelihood also over
the choice of all possible $\omega$-matrices as parameters. To remove
degeneracies, we absorbe the Lagrange multipliers $\lambda_\mu$ into
the PSSM $\omega^\mu$, and introduce 
\begin{equation}
\label{eq:xi}
\xi_i^\mu(a) = \sqrt{\lambda_\mu}\, \omega_i^\mu(a), \ \ \ i=1,...,L;\ 
\mu=1,...,p; \ a\in{\cal A}\ .
\end{equation}
The model in Eq.~(\ref{eq:maxent_hp}) thus slightly simplifies into
\begin{equation}
\label{eq:maxent_hp2} 
P(\underline A\, |\, 
\{h_i(a), \xi_i^\mu(a)\}) = \frac 1Z \exp\left\{
\sum_{\mu=1}^p \sum_{i,j=1}^L \xi_i^\mu(A_i)\xi_j^\mu(A_j) +
\sum_{i=1}^L h_i(A_i)
\right\}\ ,
\end{equation}
with parameters, which have to be estimated by maximum likelihood:
\begin{equation}
\{\hat h_i(a),\hat \xi_i^\mu(a)\}  = {\rm argmax}_{\{h_i(a), \xi_i^\mu(a)\} }
\sum_{m=1}^M \log P(\underline A^m\, |\, 
\{ h_i(a), \xi_i^\mu(a)\} )\ .
\label{eq:ml}
\end{equation}
Our model becomes therefore the standard {\em Hopfield-Potts model},
which has been introduced in \cite{cocco2013principal} in a mean-field
treatment, and the sequence motifs equal, up to the rescaling in
Eq.~(\ref{eq:xi}), the patterns in the Hopfield-Potts model.

The mean-field treatment of \cite{cocco2013principal} has both
advantages and  disadvantages with respect to our present work: on one
hand, the largely analytical mean-field solution allows
to relate the Hopfield-Potts patterns $\xi^\mu$ to the eigenvectors of
the Pearson-correlation matrix of the MSA, and their likelihood
contributions to a function of the corresponding eigenvalues. This is,
in particular, interesting since not only the eigenvectors
corresponding to large eigenvalues were found to contribute -- as one
might expect from the apparent similarity to principal-component
analysis (PCA) -- but also the smallest eigenvalues lead to large
likelihood contributions. However, the mean-field treatment leads to
a non-generative model, which does not even reproduce precisely the
single-position frequencies $f_i(a)$. The aim of this paper is to
re-establish the generative character of the Hopfield-Potts model by
more accurate interefence schemes, without loosing too much of the
interpretability of the mean-field approximation.

The model in Eq.~(\ref{eq:maxent_hp2}) contains now an exponent, which
is non-linear in the parameters $\xi^\mu$. As a consequence, the
likelihood is not convex any more, and possibly many local likelihood
maxima exist. This is also reflected by the fact that any
$p$-dimensional orthogonal transformation of the $\xi^\mu$ leaves the
probability distribution $P(\underline A)$ invariant, thus leading to
an equivalent model.

\section{Inference and interpretation of Hopfield-Potts models} 
\label{sec:inference}

\subsection{The Hopfield-Potts model as a Restricted Boltzmann
  Machine} 

\begin{figure}[htb!]
          \centering
          \includegraphics[width=0.8\textwidth]{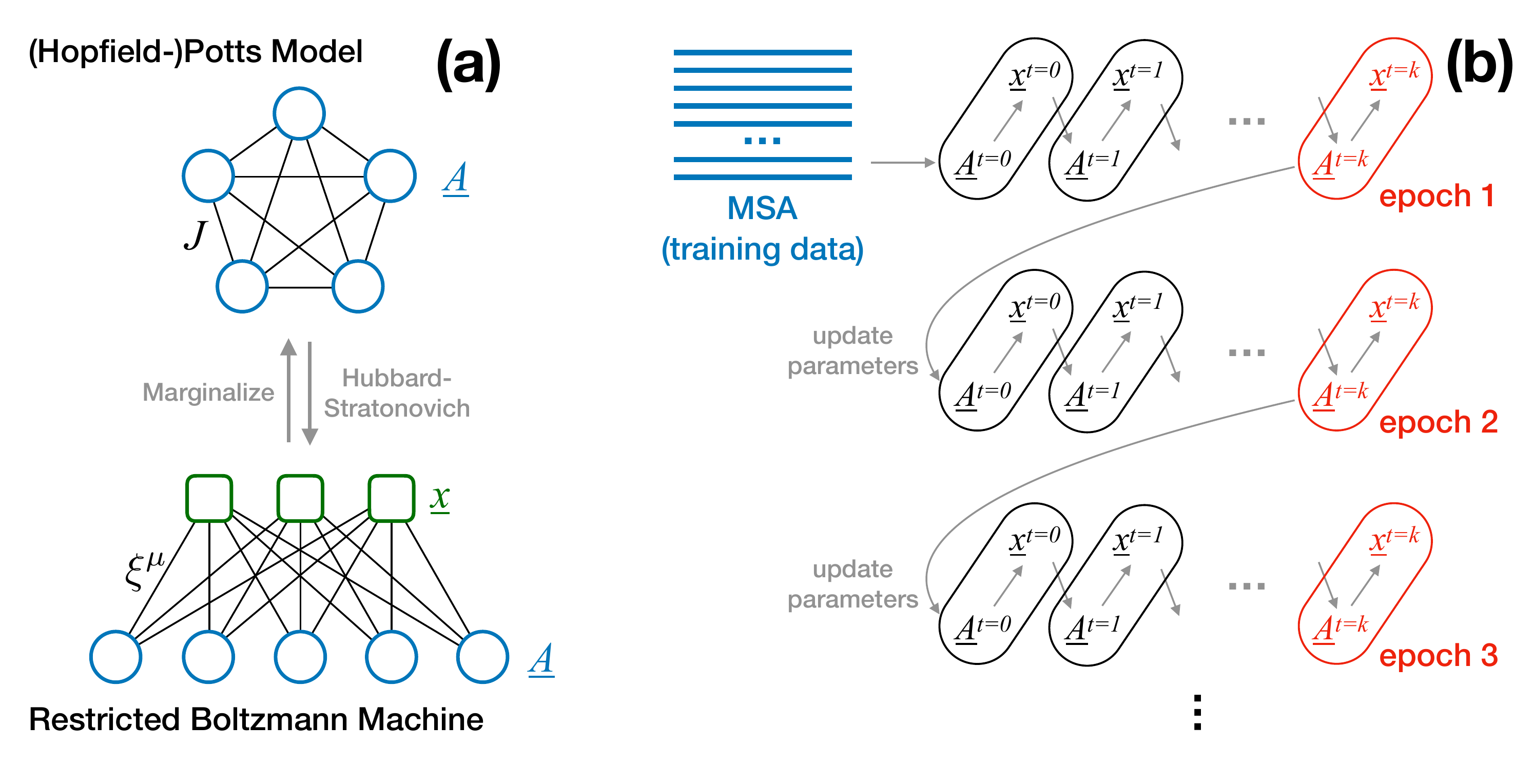}
          \caption{{\bf Panel (a)} represents the (Hopfield-)Potts model
            as a statistical model for sequences $\underline A\in {\cal
              A}^L$, typically characterized by a fully connected
            coupling matrix $J$ and local fields $h$ (not
            represented). The model can be transformed into a Restricted
            Boltzmann Machine (RBM) by introducing Gaussian hidden
            variables 
            $\underline x\in \mathbb{R}^p$, with $p$ being the rank
            of $J$. Note the bipartite graphical structure of RBM,
            which causes the conditional probabilities $P(\underline
            A\,|\,\underline x)$ and $P(\underline x\,|\,\underline A)$ to
            factorize.
            {\bf Panel (b)} shows a schematic representation of
            Persistent Contrastive Divergence (PCD). Initially the
            sample is initialized in the training data (the MSA of
            natural sequences), and then $k$ alternating steps of
            sampling from  $P(\underline A\,|\,\underline x)$
            resp. $P(\underline x\,|\,\underline A)$ are
            performed. Parameters are updated after these $k$ sampling
            steps, and sampling is continued using the updated
            parameters. 
          }
         \label{fig:rbm_pcd}
\end{figure}

The question how many and which patterns are needed for generative
modeling therefore cannot be answered properly within the mean-field 
approach. We therefore propose a more accurate inference scheme based
on {\em Restricted Boltzmann Machine} (RBM) learning
\cite{smolensky1986information,hinton2006reducing}, exploiting an
equivalence between Hopfield models and RBM originally shown in
\cite{barra2012equivalence}. To this aim, we first perform $p$
Hubbard-Stratonovich transformations to linearize the exponential in
the $\xi^\mu$, 
\begin{equation}
\label{eq:hp_to_rbm}
P(\underline A) = \frac 1{\tilde Z} \int_{\mathbb{R}^p}
\prod_{\mu=1}^p dx^\mu \exp\left\{
\sum_{i,\mu}x^\mu \xi_i^\mu(A_i) + \sum_{i} h_i(A_i) - \frac 12
\sum_\mu (x^\mu)^2
\right\}
\end{equation}
with $\tilde Z$ containing the normalizations both of the Gaussian
integrals over the new variables $x^\mu$, and the partition function
of Eq.~(\ref{eq:maxent_hp}). The distribution $P(\underline A)$ can
thus be understood as a marginal distribution of 
\begin{equation}
\label{eq:rbm}
P(\underline A, \underline x) = \frac 1{\tilde Z} \exp\left\{
\sum_{i,\mu}x^\mu \xi_i^\mu(A_i) + \sum_{i} h_i(A_i) - \frac 12
\sum_\mu (x^\mu)^2 \right\}\ ,
\end{equation}
which depends on the so-called {\it visible variables} $\underline A =
(A_1,...,A_L)$ and the {\it hidden (or latent) variables} $\underline x =
(x^1,...,x^p)$. It takes the form or a particular RBM, with a
quadratic confining
potential for the $x^\mu$: The important point is that couplings in
the RBM form a bipartite graph between visible and hidden
variables, cf.~Fig.~\ref{fig:rbm_pcd}. RBM may have more general
potentials $u_\mu(x^\mu)$ confining the values of the new random
variables $x^\mu$. This fact has been exploited in
\cite{tubiana2019learning} to cope with the limited number of
sequences in the training MSA. However, in our work we stick to
quadratic potentials in order to keep the equivalence to Hopfield-Potts
models, and thus the interpretability of patterns in terms of pairwise
residue-residue couplings via Eq.~(\ref{eq:maxent_hp2}).

\subsection{Parameter learning by persistent contrastive divergence}

Maximizing the likelihood with respect to the parameters leads, for
our RBM model, to the stationarity equations
\begin{eqnarray}
\label{eq:stationarity}
\frac 1M \sum_{m}  \delta_{A_i^m,a} &=& \langle \delta_{A_i,a}
                                        \rangle_{P(\underline A,
                                        \underline x)} \nonumber \\
\frac 1M \sum_{m}  \delta_{A_i^m,a} \langle x^\mu \rangle_ {P(\underline x\,|\,
                                        \underline A^m)}
&=& \langle \delta_{A_i,a} x^\mu
                                        \rangle_{P(\underline A,
                                        \underline x)} 
\end{eqnarray}
for all $i,a$ and $\mu$; the difference of both sides equals the
gradient of the likelihood in direction of the corresponding parameter.
While the first line matches the standard
MaxEnt form -- sample and ensemble average of an observable have to
coincide, the second line contains a mixed sample-ensemble average on
its left-hand side. Since the variables $x^\mu$ are latent and thus
not contained in the MSA, an average over their probability 
$P(\underline x\,|\,\underline A^m)$ conditioned to the sequences
$\underline A^m$ in the MSA has to be taken. Having a $P$-dependence
on both sides of Eqs.~(\ref{eq:stationarity}) is yet another
expression of the non-convexity of the likelihood function.

Model parameters $h_i(a)$ and $\xi_i^\mu(a)$ have to be fitted to
satisfy the stationarity conditions Eq.~(\ref{eq:stationarity}). This
can be done iteratively: starting from arbitrarily initialized model
parameters, we determine the difference between the left- and
right-hand sides of this equation, and use this difference to update
parameters ({\em i.e.} we perform gradient ascent of the likelihood);
each of these update steps is called an {\em epoch} of learning. A
major problem is that the exact calculation of averages over the
$(L+p)$-dimensional probability distribution $P$
is computationally infeasible. It is possible to estimate these
averages by Markov Chain Monte Carlo (MCMC) sampling, but efficient
implementations are needed since accurate parameter learning requires
in practice thousands of epochs. To this aim, we exploit the bipartite
structure of RBM: both conditional probabilities $P(\underline
A\,|\,\underline x)$ and $P(\underline x\,|\,\underline A)$ are
factorized. This allows us to initialise MCMC runs in natural
sequences from the MSA and to sample the $\underline x$ and the
$\underline A$ in alternating fashion. As a second simplification we
use {\em Persistent Contrastive Divergence} (PCD)
\cite{hinton2012practical}. Only in the first 
epoch the visible variables are initialised in the MSA sequences, and
each epoch performs only a finite number of sampling steps ($k$ for
PCD-$k$),  cf.~Fig.~\ref{fig:rbm_pcd}.B. Trajectories are continued in a
new epoch after parameter updates. If the resulting parameter changes
become small enough, PCD will thereby generate close-to-equilibrium
sequences, which form an (almost) {\em i.i.d.} sample of $P(\underline A,
\underline x)$ uncorrelated from the training set used for
initialization.

Details of the algorithm, and comparison to the simpler contrastive
divergence are given in the Appendix. Further technical details, like
regularization, are also delegated to the Appendix.

\subsection{Determining the likelihood contribution of single
  Hopfield-Potts patterns}
\label{sec:likelihood}

It is obvious, that the total likelihood grows monotonously when
increasing the number $p$ of patterns $\xi^\mu$. It is therefore
important to develop criteria, which tell us if patterns are more or
less important for modeling the protein family. To this aim we
estimate the contribution of single patterns to the likelihood, by
comparing the full model with a model, where a single pattern
$\xi^\mu$ has been removed, while the other $p-1$ patterns  and the
local fields have been retained. The corresponding normalized change
in log-likelihood reads 
\begin{equation}
  \Delta \ell_\mu = \frac 1M \sum_{m=1}^M
   \left[ \log P(\underline A^m)
    - \log P_{-\mu}(\underline A^m) 
    \right]\ ,
  \label{eq:lmu}
\end{equation}
where $P_{-\mu}$ has the same form as given in Eq.~(\ref{eq:maxent_hp}) for
$P$, but with pattern $\xi^\mu = \{\xi_i^\mu(a); i=1,...,L,\,a\in
{\cal A}\}$ removed. Plugging Eq.~(\ref{eq:maxent_hp}) into
Eq.~(\ref{eq:lmu}) , we find 
\begin{equation}
  \Delta \ell_\mu = \frac 1M \sum_{m=1}^M \left[
    \sum_{i=1}^L\xi_i^\mu(A_i^m) \right]^2+ \log \frac{Z_{-\mu}}Z \ .
  \label{eq:lmu2}
\end{equation}
The likelihood difference depends thus on the ratio of the two
partition functions $Z$ and $Z_{-\mu}$. While each of them is
individually intractable due to the exponential sum over $q^L$
sequences, the ratio can be estimated efficiently using importance
sampling. We write:
\begin{eqnarray}
  \frac{Z_{-\mu}}Z &=& \frac 1Z \sum_{\underline A\in {\cal A}^L}
   \exp\left\{ \sum_{\nu\neq \mu} \sum_{i,j=1}^L
                            \xi_i^\nu(A_i)\xi_j^\nu(A_j) +\sum_{i=1}^L
                            h_i(A_i)
                            \right\}
                            \nonumber\\
                        &=& \sum_{\underline A\in {\cal A}^L} P(\underline A)
                            \exp\left\{ -\sum_{i,j=1}^L
                            \xi_i^\mu(A_i)\xi_j^\mu(A_j) \right\} \ .
  \label{eq:importance_sampling}
\end{eqnarray}
The last expression contains the average of an exponential quantity
over $P(\underline A)$, so estimating the average by MCMC sampling of
$P$ might appear a risky idea. However, since $P$ and $P_{-\mu}$
differ only in one of the $p$ patterns, the distributions are expected
to overlap strongly, and sufficiently large samples drawn from
$P(\underline A)$ can be used to estimate $Z_{-\mu}/Z$. Note that
sampling is done from $P$, so the likelihood-contributions of all
patterns can be estimated in parallel using a single large sample of
the full model. 

Once these likelihood contributions are estimated, we can sort them,
and identify and interpret the patterns of largest importance in our
Hopfield-Potts model.

\section{Hopfield-Potts models of protein families}
\label{sec:proteins}

To understand the performance of Hopfield-Potts models in the case of
protein families, we have analysed three protein families extracted
from the Pfam database \cite{finn2016pfam}: the Kunitz/Bovine
pancreatic trypsin 
inhibitor domain (PF00014), the Response regulator receiver domain
(PF00072) and the RNA recognition motif (PF00076). They have been
selected since they have been used in DCA studies before, in our case
RBM results will be compared to the ones of bmDCA, {\em i.e.}~the generative
version of DCA based on Boltzmann Machine Learning
\cite{figliuzzi2018pairwise}. MSA are 
downloaded from the Pfam database \cite{finn2016pfam}, and sequences
with more than 5 consecutive gaps are removed; cf.~App.~\ref{app:cd}
for a discussion of the convergence problems of PCD-based inference in
case of extended gap stretches. The resulting MSA dimensions for the
three families are, in the order given before, $L=52 / 112 / 70$ and
$M = 10 657 / 15 000 / 10 000$. As can be noted, the last two MSA have
been subsampled randomly since they were  very large, and the running
time of the PCD algorithm is linear in the sample size. The MSA for
PF00072 was chosen to be slightly larger because of the longer
sequences in this family.

In the following sections, results are described in detail for the
PF00072 response regulator family. The results for the other protein
families are coherent with the discussion; they are moved to the
App.~\ref{app:cd} for the seek of conciseness of our presentation.

\subsection{Generative properties of Hopfield-Potts models}

PCD is able, for all values of the pattern number $p$, to reach
parameter values satisfying the
stationarity conditions Eqs.~(\ref{eq:stationarity}). This is not only
true when these are evaluated using the PCD sample propagated via
learning from epoch to epoch, but also when the inferred model is
resampled using MCMC, i.e.~when the right-hand side of
Eqs.~(\ref{eq:stationarity}) is evaluated using an {\em i.i.d.} sample of
the RBM.

\begin{figure}[htb!]
          \centering
          \includegraphics[width= 0.8\textwidth]{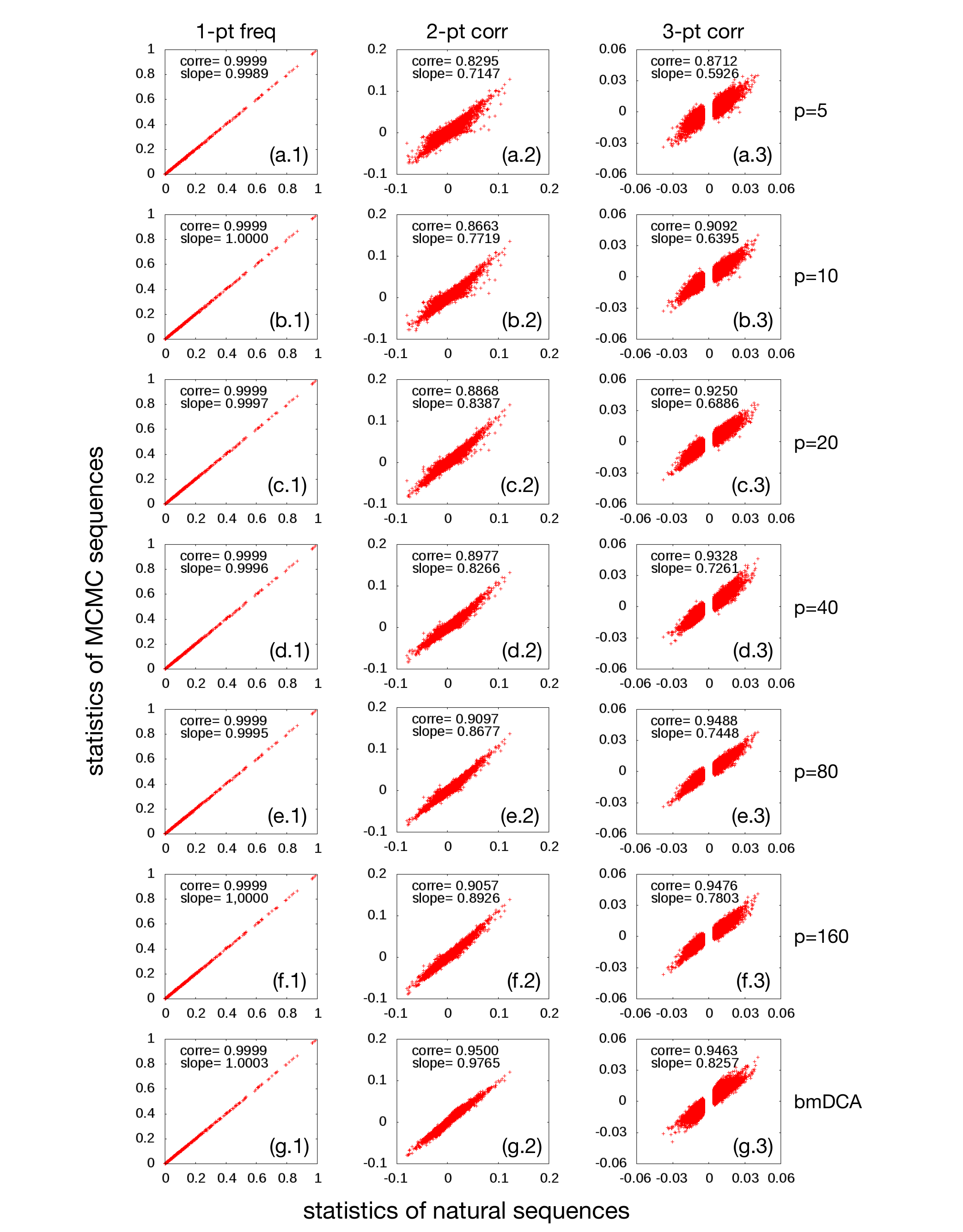}
          \caption{Statistics of natural sequences (PF00072,
            horizontal axes) vs. MCMC samples (vertical axes) of 
            Hopfield-Potts models for values of
            $p\in\{5,10,20,40,80,160\}$ and for a full-rank Potts model
            inferred using bmDCA. The first column (Panels (a.1-g.1))
            shows the 1-point frequencies $f_i(a)$ for all pairs
            $(i,a)$ of sites and 
            amino-acids, the other two column show the connected 2-
            and 3-point functions $c_{ij}(a,b)$ (Panels (a.2-g.2)) and
            $c_{ijk}(a,b,c)$ (Panels (a.3-g.3)). Due to the huge
            number of combinations 
            for the three-point correlations, only the 100,000 largest
            values (evaluated in the training MSA) are shown. The
            Pearson correlations and the slope of the best linear fit
            are inserted in each of the panels.
          }
         \label{fig:123pointcorrs}
\end{figure}

In the leftmost column of Fig.~\ref{fig:123pointcorrs} (Panels
(a.1-g.1)), this is shown for the single-site frequencies, {\it
  i.e.}~for the first of Eqs.~(\ref{eq:stationarity}). The horizontal
axis shows the statistics extracted from the original data collected
in the MSA, while the vertical axis measures the same quantity in an
{\em i.i.d.} sample extracted from the  inferred model $P(\underline
A,\underline x)$. The fitting quality is comparable to the one
obtained by bmDCA, as can be seen by comparison with the last panel in
the first column of Fig.~\ref{fig:123pointcorrs}. 

The other two columns of the figure concern the {\it generative}
properties of RBM: connected two-point correlations (Panels (a.2-g.2))
and three-point correlations (Panels (a.3-g.3)),
\begin{eqnarray}
  \label{eq:2-3-pt-corr}
  c_{ij}(a,b) &=& f_{ij}(a,b) - f_{i}(a) f_{j}(b)\ , \nonumber\\
  c_{ijk}(a,b,c) &=& f_{ijk}(a,b,c) - f_{ij}(a,b) f_{k}(c) -
                     f_{ik}(a,c) f_{j}(b) - f_{jk}(b,c) f_{i}(a)
                     +2 f_{i}(a) f_{j}(b) f_{k}(c)\ ,
\end{eqnarray}
with the three-point frequencies $f_{ijk}(a,b,c)$ defined in analogy
to Eqs.~(\ref{eq:fi},\ref{eq:fij}). Note that in difference to DCA,
already the two-point correlations are not fitted directly by the RBM,
but only the second moments related to the Hopfield-Potts patterns. This
becomes immediately obvious for the case $p=0$, where RBM reduce to
simple profile models of statistically independent sites, but remains
true for all values of $p<(q-1)L$. Note also that connected
correlations are used, since the frequencies $f_{ij}$ and $f_{ijk}$
contain information about the fitted $f_i$, and therefore show
stronger agreement between data and model. 

The performance of RBM is found to be, up to statistical fluctuations,
monotonous in the pattern number $p$. As in the mean-field
approximation \cite{cocco2013principal}, no evident overfitting
effects are observed. 
Even if not fitted explicitly, as few as $p=20-40$ patterns are
sufficient to faithfully reproduce even the non-fitted two- and
three-point correlations. This is very astonishing, since only about
1.7-3.5\% of the parameters of the full DCA model are used: the $p$
patterns are given by $p(q-1)L$ parameters, while DCA has
$(q-1)^2{L\choose 2}$ independently inferred couplings. The times
needed for accurate inference decrease accordingly: in some cases, a
slight decrease in accuracy of bmDCA is observed as compared to RBM
with the largest $p$; this could be overcome by iterating the inference
procedure for further epochs. 

\subsection{Strong couplings and contact prediction}

One of the main applications of DCA is the prediction of contacts
between residues in the three-dimensional protein fold, based only on
the statistics of homologous sequences. To this aim, we follow
\cite{ekeberg2013improved} and translate
$q\times q$ coupling matrices $J_{ij}(a,b)=\sum_{\mu=1}^p \xi_i^\mu(a)
\xi_j^\mu(b)$ for individual site pairs $(i,j)$ into
scalar numbers by first calculating their Frobenius norm,
\begin{equation}
F_{ij} = F_{ji}= \sum_{a,b\in{\cal A}} J_{ij}(a,b)^2,
\end{equation}
followed by the empirical average-product correction (APC)
\begin{equation}
F_{ij}^{APC} = F_{ij} - \frac{F_{i\cdot}F_{\cdot j}}{F_{\cdot\cdot}}\ ,
\end{equation}
where the $\cdot$ denotes an average over the corresponding index:
\begin{eqnarray}
  F_{i\cdot} &=& \frac 1{L-1} \sum_{k} F_{ik} \nonumber\\
  F_{\cdot j} &=& \frac 1{L-1} \sum_{k} F_{kj} \\
  F_{\cdot\cdot} &=& \frac 2{L(L-1)} \sum_{k<l} F_{kl} \nonumber
\end{eqnarray}
The APC is intended to remove systematic non-functional bias due to
conservation and phylogeny. These quantities are sorted, and the
largest ones are expected to be contacts.

\begin{figure}[htb!]
          \centering
          \includegraphics[width= 0.9\textwidth]{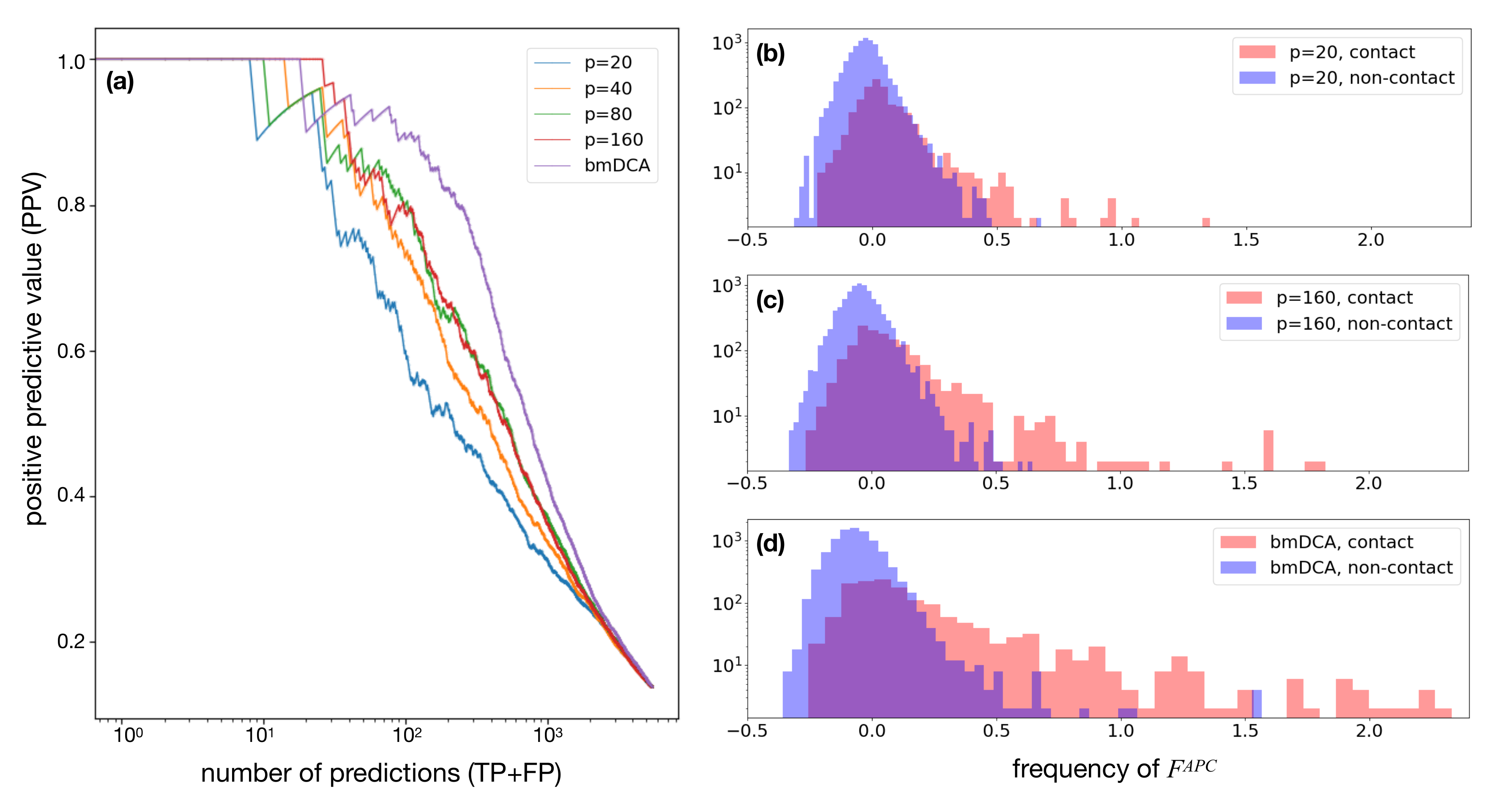}
          \caption{Panel (a) shows the positive predictive value
            (PPV) for contact prediction as a function of the number
            of predictions, for various values of the pattern number
            $p$ and for bmDCA. Panels (b-d) show, for $p=20,\ 160$
            and bmDCA, the distribution of coupling scores
            $F_{ij}^{APC}$. All residue pairs are grouped into
            contacts (red) and non-contacts (blue). The best contact
            predictions correspond to the positive tail of the red
            histogram, which becomes more pronounced when increasing
            $p$ or even going to bmDCA.
          }
         \label{fig:ppv}
\end{figure}

The results for several values of $p$ and for bmDCA are depicted in
Fig.~\ref{fig:ppv}.a: the positive predictive value (PPV) is the fraction of
true positives (TP) among the first $n$ predictions, as a function of
$n$. TP are defined as native contacts in a reference protein
structure (PDB ID 3ilh \cite{ibrahim2009pdb} for PF00072), with a
distance cutoff of  $8{\rm \AA}$ between the closest pair of heavy
atoms forming each residue. Pairs in vicinity along the peptide chain
are not considered in this prediction, since they are trivially in
contact: in coherence with the literature standard,
Fig.~\ref{fig:ppv} only considers predictions with $|i-j|\geq 5$.

Despite the fact, that even for as few as $p=20-40$ patterns the model
appears to be generative, {\em i.e.}~non-fitted statistical observables are
reproduced with good accuracy, the PPV curves depend strongly on the
pattern number $p$. Up to statistically probably insignificant
exceptions, we observe a monotonous dependence on $p$, and none of the
RBM-related curves reaches the performance of the full-rank $J_{ij}$
matrices of bmDCA. Even large values of $p$, where RBM have more than
30\% of the parameters of the full Potts model, show a drop in
performance in contact prediction.

Can we understand this apparent contradiction: similarly accurate
reproduction of the statistics, but reduced performance in contact
prediction? To this end, we consider in Figs.~\ref{fig:ppv}.b-d the
histograms of coupling strengths $F_{ij}^{APC}$ divided into two
subpopulations: values for sites $i,j$ in contact are represented by
red, for distant sites by blue histograms. It becomes evident
that the rather compact histogram of non-contacts remains almost
invariable with $p$ (even if individual coupling values do change),
but the histogram of contacts changes systematically: the tail of
large $F_{ij}^{APC}$ going beyond the upper edge of the blue histogram
is less pronounced for small $p$. However, in the procedure described
before, these $F_{ij}^{APC}$-values provide the first contact
predictions. 

The reduced capacity to detect contacts for small $p$ is related to
the properties of the Hopfield-Potts model in itself.  While the
esidue-residue contacts form a sparse graph, the
Hopfield-Potts model is explicitly constructed to have a low-rank
coupling matrix $(J_{ij}(a,b))$. It is, however, hard to represent a
generic sparse matrix by a limited number of possibly distributed
patterns. Hopfield-Potts models are more likely to detect distributed
sequence signals than localized sparse ones. However, for larger
pattern numbers $p$ we are able to detect more and more localized
signals, thereby improving the contact prediction, until BM and
Hopfield-Potts models become equivalent for $p=(q-1)L$.

This observation establishes an important limitation to the generative
character of Hopfield-Potts models with limited pattern numbers:  The
applicability of DCA for residue-residue contact prediction has
demonstrated that physical contacts in the three-dimensional structure
of proteins introduce important constraints on sequence evolution. A
perfectly generative model should respect these constraints, and thus
lead to a contact prediction being at least as good as the one
obtained by full DCA, cf.~also the discussion in the outlook of this
article. 

\subsection{Likelihood contribution and interpretation of selected 
  sequence motifs}

\begin{figure}[htb!]
          \centering
          \includegraphics[width= 0.6\textwidth]{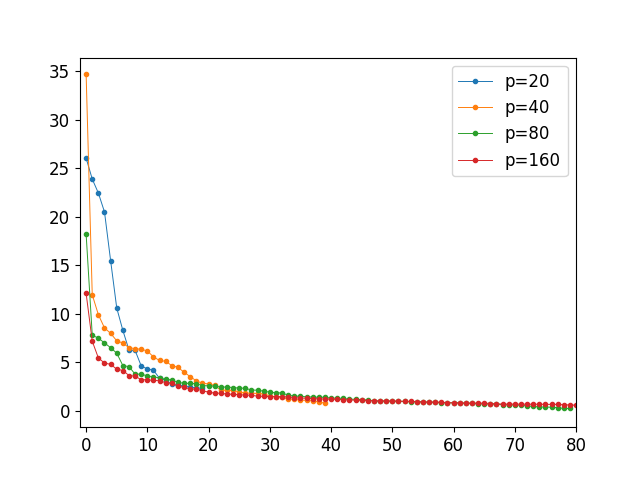}
          \caption{Likelihood contribution of the individual patterns,
            for pattern numbers $p=20,40,80,160$.
          }
         \label{fig:ll_pattern}
\end{figure}

So what do the patterns represent? In Sec.~\ref{sec:likelihood}, we
have discussed how to estimate the likelihood contribution of
patterns, thereby being able to select the most important patterns in
our model. Fig.~\ref{fig:ll_pattern} displays the ordered
contributions for different values of $p$. We observe that, for small
$p$, the distribution becomes more peaked, with few patterns having
very large likelihood contributions. For larger $p$, the contributions
are more distributed over many patterns, which collectively represent
the statistical features of the data set.

\begin{figure}[htb!]
          \centering
          \includegraphics[width= 0.9\textwidth]{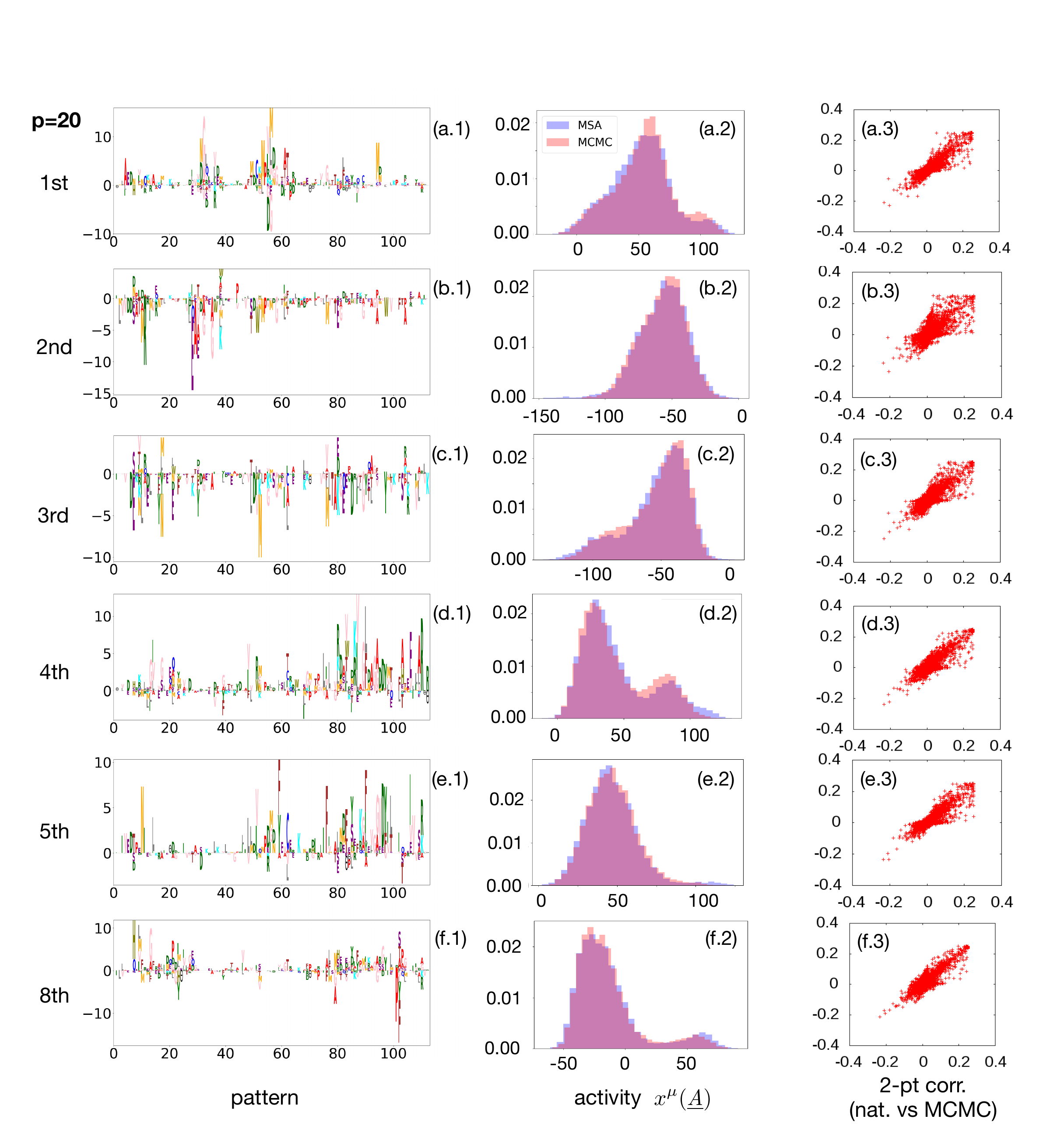}
          \caption{The five patterns of highest likelihood-contribution
            for $p=20$, the 8th ranking is added since used later in
            the text. The left Panels (a.1-f.1) show the patterns in logo
            representation, the letter-size is given by the corresponding
            element $\xi_i(a)$. The middle panels (a.1-f.2) show
            the distribution of the activities, {\it i.e.}~the projections
            of sequences onto the patterns. The blue histogram
            contains the natural sequences from the training MSA, the
            red histogram sequences sampled by MCMC from the
            Hopfield-Potts model. The right-hand side (Panels
            (a.3-f.3)) shows the connected 2-point correlations of the
            natural data (horizontal axis) vs.~data sampled from
            $P_{-\mu}(\underline A)$, i.e.~a Hopfield-Potts model with
            one pattern removed. Strong deviations from the diagonal
            are evident.
          }
         \label{fig:first_patterns}
\end{figure}

Fig.~\ref{fig:first_patterns} represents the first five patterns for
$p=20$. Panels (a.1-e.1) represent the pattern $\xi_i^\mu(a)$ as a
sequence logo, a standard representation in sequence
bioinformatics. Each site $i$ corresponds to one position, the
possible amino-acids are shown by their one-letter code, the size of
the letter being proportional to $|\xi_i^\mu(a)|$, according to the sign
of $\xi_i^\mu(a)$, letters are represented above or below the zero
line. The alignment gap is represented as a minus sign in an oval
shape, which allows to represent its size in the current pattern.

Patterns are very distributed, both in terms of the sites and
amino-acids with relatively large entries $\xi_i(a)$. This makes a
direct interpretation of patterns  without prior knowledge rather
complicated. The distributed nature of patterns explains also why they
are not optimal in defining localised contact predictions. Rather than
identifying contacting residue pairs, the patterns define larger
groups of sites, which are connected via a dense network of comparable
couplings. However, as we will see in the next section, the sites of
large entries in a pattern define functional regions of proteins,
which are important in sub-ensembles of proteins of strong (positive
or negative) activity values along the pattern under consideration. In
particular, we will show that the largest entries may have an
interpetation connecting structure and function to sequence in protein 
sub-families. 

The middle column (Panels (a.2-e.2)) shows a histogram of
pattern-specific activities of single sequences, {\em i.e.}~of
\begin{equation}
  x^\mu(\underline A) = \sum_{i=1}^N \xi_i^\mu(A_i)\ .
\end{equation}
Note that, up to the rescaling in Eq.~(\ref{eq:xi}), these numbers
coincide with the sequence motifs, introduced in Eq.~(\ref{eq:motif})
at the beginning of this article. They also equal the average value of
the latent variable $x^\mu$ given sequence $\underline A$. The blue 
histograms result from the natural sequences collected in the 
training MSA. They coincide well with the red histograms, which are
calculated from an {\em i.i.d.}~MCMC sample of our Hopfield-Potts
model, including the bimodal structure of several histograms. This is
quite remarkable: the Hopfield-Potts model was derived, in the
beginning of this work, as the maximum-entropy model reproducing the
first two moments of the activities $\{ x^\mu(\underline A^m)
\}_{m=1...M}$. Finding higher-order features like bimodality is again
an expression of the generative power of Hopfield-Potts models.

Figs.~\ref{fig:first_patterns}.a.3-\ref{fig:first_patterns}.e.3 prove the
importance of individual patterns for the inferred model. The panels
show the two-point correlations $c_{ij}(a,b)$ of the natural data
(horizontal axis) vs. the one of samples drawn from the distributions 
$P_{-\mu}(\underline A)$, introduced in
Eq.~(\ref{eq:lmu}) as Hopfield-Potts models of $p-1$ patterns, with
pattern $\xi^\mu$ removed (vertical axis). The coherence of the
correlations is strongly reduced when compared to the full model,
which was shown in Fig.~\ref{fig:123pointcorrs}: removal even of a
single pattern has a strong global impact on the model statistics.

\subsection{Sequence clustering}

As already mentioned, some patterns show a clear bimodal activity
distribution, {\em i.e.}~they identify two statistically distinct subgroups
of sequences. The number of subgroups can be augmented by using more  
than one pattern, {\em i.e.}~combinations of patterns can be used to
cluster sequences.

To this aim, we have selected three patterns (number 6, 13 and 14)
with a pronounced bimodal structure from the model with $p=20$
patterns. In terms of likelihood contribution, they have ranks 8, 4
and 1 in the contributions to the log-likelihood,
cf. Fig.~\ref{fig:first_patterns}.

\begin{figure}[htb!]
          \centering
          \includegraphics[width=\textwidth]{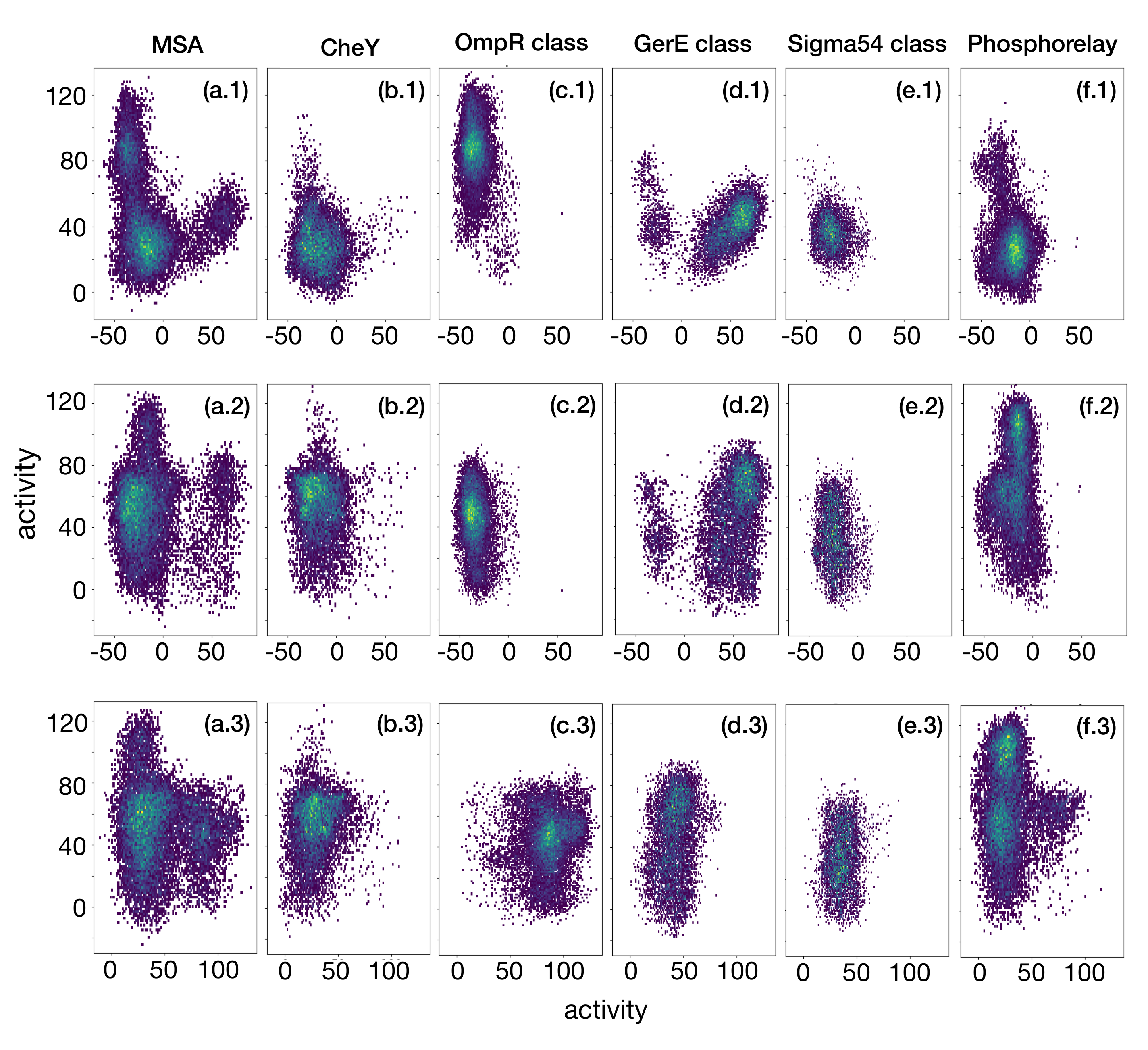}
          \caption{Patterns with multimodal activity distributions for
            the set of all MSA sequences can be used to cluster
            sequences. The rows show combinations of
            patterns 6-13 (Panels (a.1-f.1)),  6-14  (Panels
            (a.2-f.2)) and 13-14  (Panels (a.3-f.3)). Each sequence
            corresponds to a density-colored dot. A strongly clustered
            structure is clearly visible. When dividing the full MSA into
            functional subclasses, we can relate clusters to
            subclasses, and thus patterns to biological function.
          }
         \label{fig:clustering}
\end{figure}

The clustered organization of response-regulator sequences becomes
even more evident in the two-dimensional plots characterizing
simultaneously two activity distributions. The results for all pairs
of the three patterns are displayed in Fig.~\ref{fig:clustering},
Panels (a.1-a.3). As a first observation, we see that the main modes
of the activity patterns give rise to one dominant cluster. Smaller
cluster deviate from the dominant one in a single pattern, but show
compatible activities in the other patterns -- the two-dimensional
plots therefore show typically an L-shaped sequence distribution, and
three clusters, instead of the theoretically possible four
combinations of activity models. It appears that single patterns
identify the particularities of single subdominant sequence clusters.

We have chosen the response-regulator protein-domain family in this
paper also due to the fact, that it constitutes a functionally well
studied and diversified family. Response regulators are predominantly
used in bacterial signaling systems:
\begin{itemize}
\item In {\it chemotaxis}, they appear as
single-domain proteins named CheY, which transmit the signal from
kinase proteins (activated by signal reception) to flagellar motor proteins,
which trigger the movement of the bacteria. CheY proteins can be
identified in our MSA as those coming from single-domain proteins,
{\em i.e.}~with lengths compatible to the PF00072-MSA width
$L=112$. We have selected a sub-MSA consisting of all proteins with
total sequence lengths between 110 and 140 amino acids.
\item In {\it two-component signal transduction} (TCS), response regulators are
typically transcription factors, which are activated by
signal-receiving histidine sensor kinases. The corresponding proteins
contain two or three domains, in particular a DNA-binding domain,
which is actually responsible for the transcription-factor activity of
the activated response-regulator protein. According to the present
DNA-binding domain, these TCS proteins can be subdivided into
different classes, the dominant ones are the OmpR, the GerE and the
Sigma54-HTH8 classes, we identified three sub-MSA corresponding to
these classes by co-occurrence of the DNA-binding domains with the
response-regulator domain in the same protein. The different
DNA-binding domains are indicative for distinct homo-dimer structures
assumed by the active transcription factors; DCA run on the sub-MS
identifies their specific sub-family interfaces \cite{uguzzoni2017large}. 
\item {\it Phosphorelays} are similar to TCS, but consist of more
complex multi-component signaling pathways. In these systems, found in 
bacteria and plants,
response-regulator domains are typically fused to the histidine-sensor
kinases. They do not act as transcription factors, but transduct a
signal to a phosphotransferase, which finally activates a down-stream
transcription factor of the same architecture mentioned in the last
paragraph. We identified a class of response regulator domains, which
are fused to a histidine kinase domain.  In terms of domain
architecture and protein length, this subfamily is extremely
heterogenous. 
\end{itemize}
Panel columns (a-f) in Fig.~\ref{fig:clustering} show the activities
of these five sub-families. It is evident, that distinct sub-MSA fall
actually into distinct clusters according to these three patterns:
\begin{itemize}
\item The CheY-like single domain proteins (Panels (b.1-b.3)) fall,
  according to all three patterns, into the dominant mode.
\item The OmpR-class transcription factors (Panels (c.1-c.3))
show a distinct distribution of higher activities for the second of
the patterns (which actually has the most pronounced bimodal
structure, probably due to the fact that the OmpR-class forms the
largest sub-MSA). As can be seen in Fig.~\ref{fig:clustering}, this
pattern has the largest positive entries in the region
of positions 80-90 and 100-110. Interestingly, these regions define
the interface of OmpR-class transcription-factor homodimerization,
cf.~\cite{uguzzoni2017large}. In accordance with this structural
interpretation, we also find a periodic structure of period 3-4 of the
large entries in the pattern, which reflects the fact that the
interface is formed by two helices, which lead to a periodic exposure
of amino-acids in the protein surface.
\item The GerE-class (Panels (d.1-d.3)) differs in activities in
  direction of the first 
  pattern, only GerE-class proteins have positive, all other have
  negative activities. Dominant positive entries are found in regions
  5-15 and 100-105, again identifying the homo-dimerization interface,
  cf.~\cite{uguzzoni2017large}.
\item The Sigma54 class (Panels (e.1-e.3)) does not show a distinct
  distribution of 
  activities according to the three selected patterns. It is
  located together with the CheY-type sequences. However, when
  examining all patterns, we find that pattern number 5 (ranked 6th
  according to the likelihood contribution) is almost perfectly
  discriminating the two.
\item Last but not least, the response-regulators fused to histidine
  kinases in phosphorelay systems (Panels (f.1-f.3)) show a distinct
  activity distribution according to the third pattern, mixing a part
  of activities compatible with the main cluster, and others being
  substantially larger (this mixing results presumably from the
  previously mentioned heterogenous structure of this sub MSA).
  Structurally known complexes between 
  response-regulators and histidine phosphotransferases (PDB ID 4euk
  \cite{bauer2013structure}, 1bdj \cite{kato1999structure}) show the
  interface located in residues 5-15, 30-32 and 50-55, 
  regions being important in the corresponding pattern. It appears
  that the pattern selects the particular amino-acid composition of
  this interface, which is specific to the phosphorelay sub-MSA.
\end{itemize}
These observations do not only show that the patterns allow for
clustering sequences into subMSA, but the discriminating positions in
the patterns have a clear biological interpretation. This is very
interesting, since the analysis in \cite{uguzzoni2017large} required
a prior clustering of the initial MSA into sub-MSA, and the
application of DCA to the individual sub-MSA. Here we have inferred
only one Hopfield-Potts model describing the full MSA, and the
patterns automatically identify 
biologically reasonable sub-families together with the sequence
patterns characterizing them. The prior knowledge needed in
\cite{uguzzoni2017large} is not needed here; we use it only for the
posterior interpretation of the patterns.

It is also important to remember that sequence clustering can also be
obtained by a technically simpler PCA (principal-component
analysis). PCA is based on the leading eigenvectors of the
data-covariance matrix, {\em i.e.}~exclusively on the largest
eigenvalues. The potential differences were already 
discussed in \cite{cocco2013principal} in the context of the
mean-field approximation of Hopfield-Potts models. It was shown that
not only the eigenvectors with large eigenvalues lead to important
contributions in likelihood, but also those corresponding to the
smallest eigenvalues. Both tails of the spectrum are thus important
for the statistical description of protein-sequence ensembles. A
second drawback of PCA as compared to our approach is the
non-generative character of PCA. No explicite statistical model is 
learned, but the data covariance matrix is simply approximated by a
low-rank matrix. 

\section{Conclusion and outlook}
\label{sec:conclusion}

In this paper, we have rederived Hopfield-Potts models as statistical
models for protein sequences by selection of additive sequence
motifs. Statistical sequence models are required to reproduce the
first and second moments of the empirical motif distributions ({\em
  i.e.} over the MSA of natural sequences). Within a maximum-entropy
approach, these motifs are found to be (up to a scaling factor) the
Hopfield-Potts patterns defining a network of residue-residue
couplings. In addition to the maximum-entropy framework, which is
built upon known observables, the Hopfield-Potts model adds a step of
variable selection: the probability of the sequence data is maximised
over all possible selections of sequence motifs.

The quadratic coupling terms can be linearised using a
Hubbard-Stratonovich transformation. When the Gaussian variables 
introduced in this transformation are interpreted as latent random
variables, the Hopfield-Potts model takes the form of a Restricted
Boltzmann Machine. This interpretation, originally introduced in
\cite{barra2012equivalence}, allows for the application of 
efficient inference techniques, like persistent contrastive
divergence, and therefore for the accurate inference of the
Hopfield-Potts patterns for any given MSA of a homologous protein
family. 

We find that Hopfield-Potts models acquire interesting generative
properties even for a relatively small number of parameters
(p=20-40). They are able to reproduce non-fitted properties like
higher-order covariation of residues. Also the bimodality observed in
the empirical activity distributions (i.e. the projection of the natural
sequences onto individual Hopfield-Potts patterns) is not
automatically guaranteed when using only the first two moments for
model learning, but it is recovered with high accuracy in the activity
distributions of artificially sequences sampled from the model. This
observation is not only interesting in the context of generative-model
learning, but forms the basis of sequence clustering according to
interpretable sequence motifs in the main text. 

The Hopfield-Potts patterns, or sequence motifs, are typically found to
be distributed over many residues, thereby representing global
features of sequences. This observation explains, why Hopfield-Potts
models tend to loose accuracy in residue-residue contact prediction,
as compared to the full-rank Potts models normally used  in Direct
Coupling Analysis: the sparsity of the residue-residue contact network
cannot be represented easily via few distributed sequence motifs,
which describe more global patterns of sequence variability. Despite
the fact, that Hopfield-Potts models reproduce also non-fitted
statistical observables, the loss of accuracy in contact prediction
demonstrates that these models are not fully generative, and
alternative concepts for parameter reduction should be explored.

Individual sequences from the input MSA can be projected onto the
Hopfield-Potts patterns, resulting in sequence-specific activity
values. Some patterns show a mono-modal histogram for the protein
family. They introduce a dense network of relatively small couplings
between positions with sufficiently large entries in the pattern,
without dividing the family into subfamilies. These patterns have great
similarity to the concept of protein sectors, which was introduced in
\cite{halabi2009protein,rivoire2016evolution} to detect distributed
modes of sequence coevolution. However, the conservation-based
reweighting used in determining sectors is not present in the
Hopfield-Potts model, and the precise relationship between both ideas
remains to be elaborated. Other Hopfield-Potts patterns show a bimodal 
activity distributions, leading to the detection of functional
sub-families. Since these are defined by, e.g., the positive vs. the
negative entries of the pattern, the entries of large absolute value
in the patterns identify residues, which play a role similar to
so-called specificity determining residues
\cite{casari1995method,mirny2002using}, {\em i.e.}~residues, which are
conserved inside specific sub-families, but 
vary between sub-families. Both concepts -- sectors and
specificity-determining residues -- emerge naturally in the
context of Hopfield-Potts families, even if their precise mathematical 
definitions differ, and thus also their precise biological
interpretations. 

These observations open up new ways of parameter reduction in
statistical models of protein sequences: the sparsity of contacts,
which are expected to be responsible for a large part of localized
residue covariation in protein evolution, has to be combined with the
low-rank structure of Hopfield-Potts models, which detect distributed
functional sequence motifs. However, distributed patterns may also be
related to phylogenetic correlations, which are present in the data,
cf.~\cite{qin2018power}. As has been shown recently in a
heuristic way \cite{zhang2016improving},  the decomposition 
of sequence-data covariance matrices or couplings matrices into a sum
of a sparse and a low-rank matrix can substantially improve contact
prediction, if only the sparse matrix is used. 

Combining this idea with the idea of generative modeling seems a
promising road towards parsimonious sequence models, which in turn
would improve parameter interpretability and reduce overfitting
effects, both limiting factors of current versions of DCA. In this
context, it will also be interesting to explore more general
regularization strategies which favor more localized sequence motifs,
or Hopfield-Potts patterns, thereby unifying sparse and low-rank 
inference in a single framework of parameter-reduced statistical
models for biological sequence ensembles.

\section*{Acknowledgements}
We are particularly grateful to Pierre Barrat-Charlaix, Giancarlo
Croce, Carlo Lucibello, Anna-Paola Muntoni, Andrea Pagnani, Edoardo
Sarti, Jerome Tubiana and Francesco Zamponi for numerous discussions
and assistance with data.  

We acknowledge funding by the EU H2020 research and innovation 
programme MSCA-RISE-2016 under grant agreement No. 734439 InferNet. 
KS  acknowledges an Erasmus Mundus TEAM (TEAM -- Technologies for
information and communication, Europe - East Asia Mobilities)
scholarship of the European Commission in 2017/18, and a doctoral
scholarship of the Honjo International Scholarship Foundation since
2018.  

\bibliography{review} 

\appendix

\section{Results for other protein families}
\label{app:other_pfam}

The first appendix is dedicated to other protein families. As
discussed in the main text, we have analyzed three distinct families,
and discussed only one in full detail in the main text. Here we
present the major results -- generative properties, contact prediction
and selected collective variables (patterns) -- for two more
families. These results show the general applicability of our approach
beyond the specific response-regulator family used in the main text.

\subsection{Kunitz/Bovine pancreatic trypsin inhibitor domain PF00014}

\begin{figure}[htb!]
          \centering
          \includegraphics[width=0.8\textwidth]{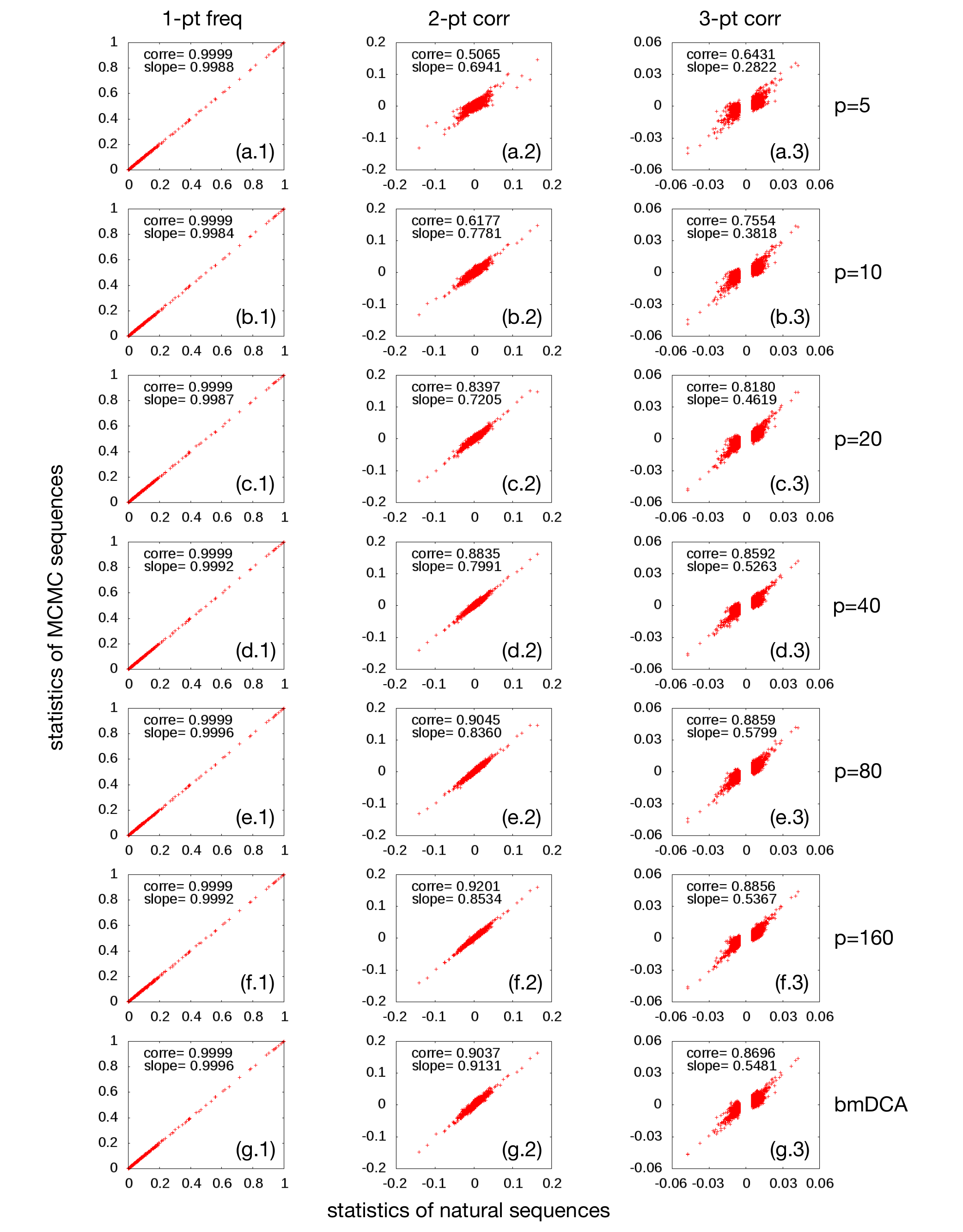}
          \caption{Same as Fig.~\ref{fig:123pointcorrs}, but for the
            protein family PF00014.
          }
         \label{fig:pf14_1}
\end{figure}

\begin{figure}[htb!]
          \centering
          \includegraphics[width=0.8\textwidth]{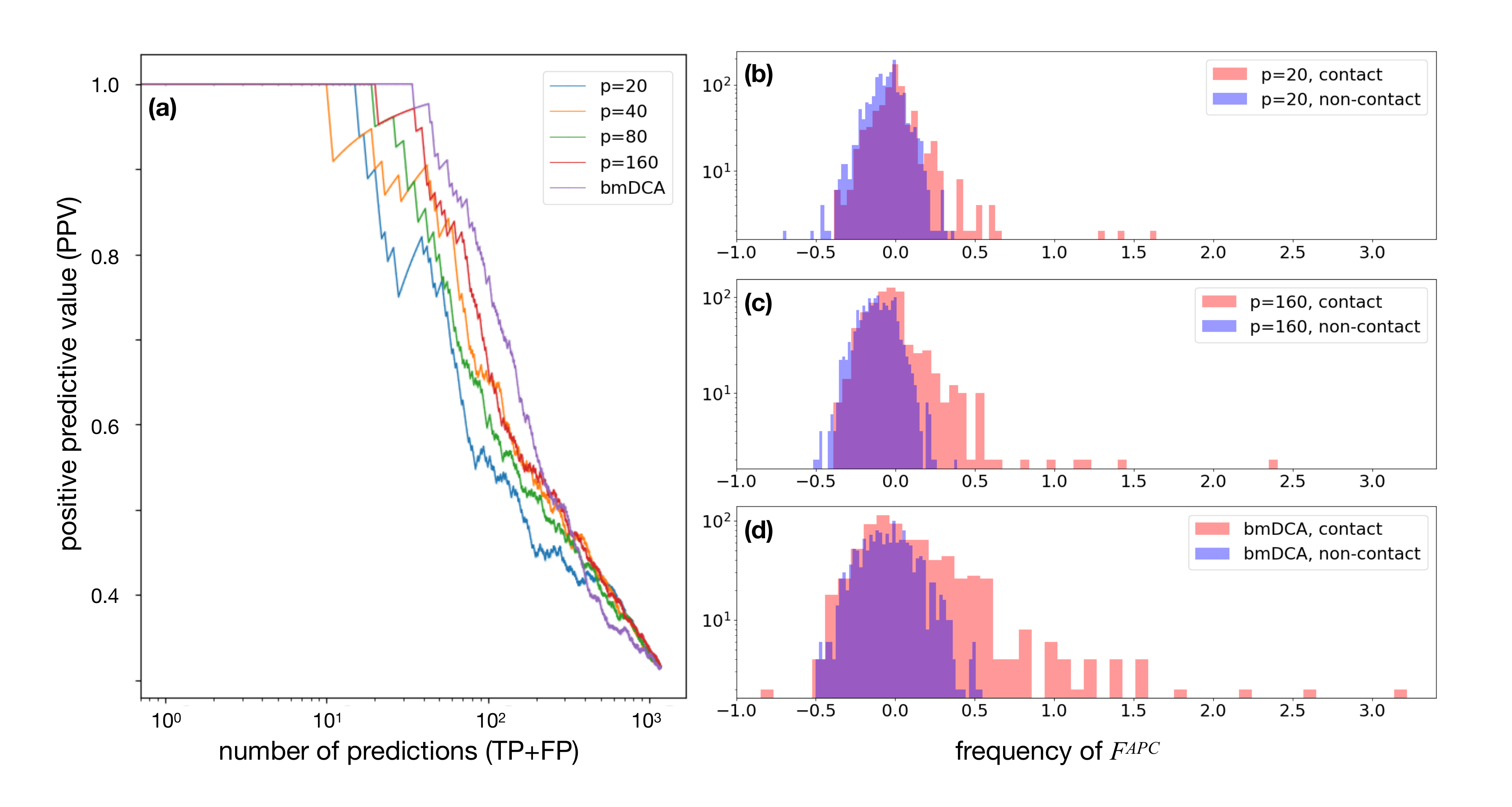}
          \caption{Same as Fig.~\ref{fig:ppv}, but for the
            protein family PF00014.
          }
         \label{fig:pf14_2}
\end{figure}

\begin{figure}[htb!]
          \centering
          \includegraphics[width=0.8\textwidth]{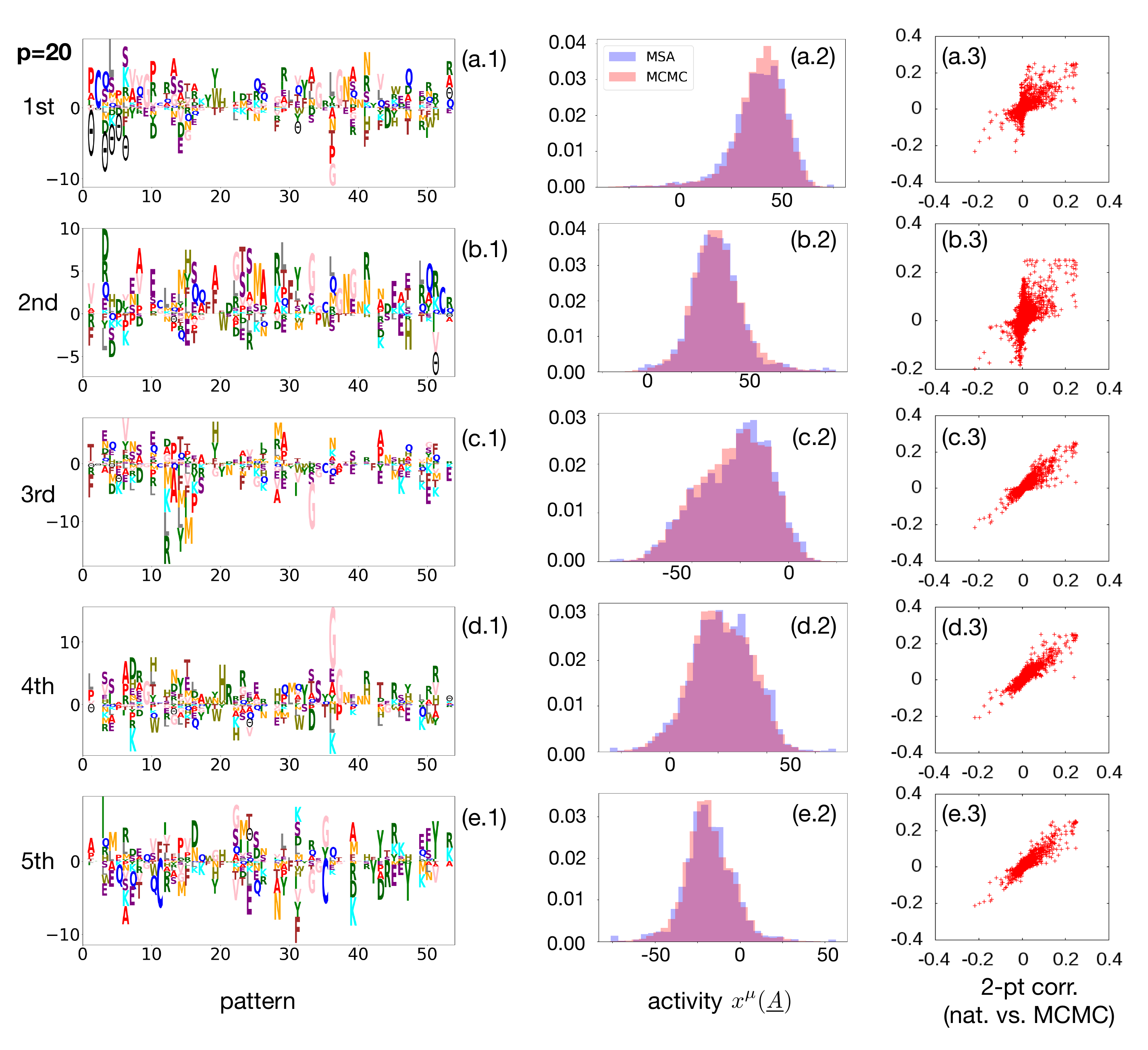}
          \caption{Same as Fig.~\ref{fig:first_patterns}, but for the
            protein family PF00014.
          }
         \label{fig:pf14_3}
\end{figure}

Figs.~\ref{fig:pf14_1}, \ref{fig:pf14_2} et \ref{fig:pf14_3} display
the major results for the PF00014 protein family. PPV curves are
calculated using PDB ID 5pti \cite{wlodawer1984structure}.

\subsection{RNA recognition motif PF00076}

\begin{figure}[htb!]
          \centering
          \includegraphics[width=0.8\textwidth]{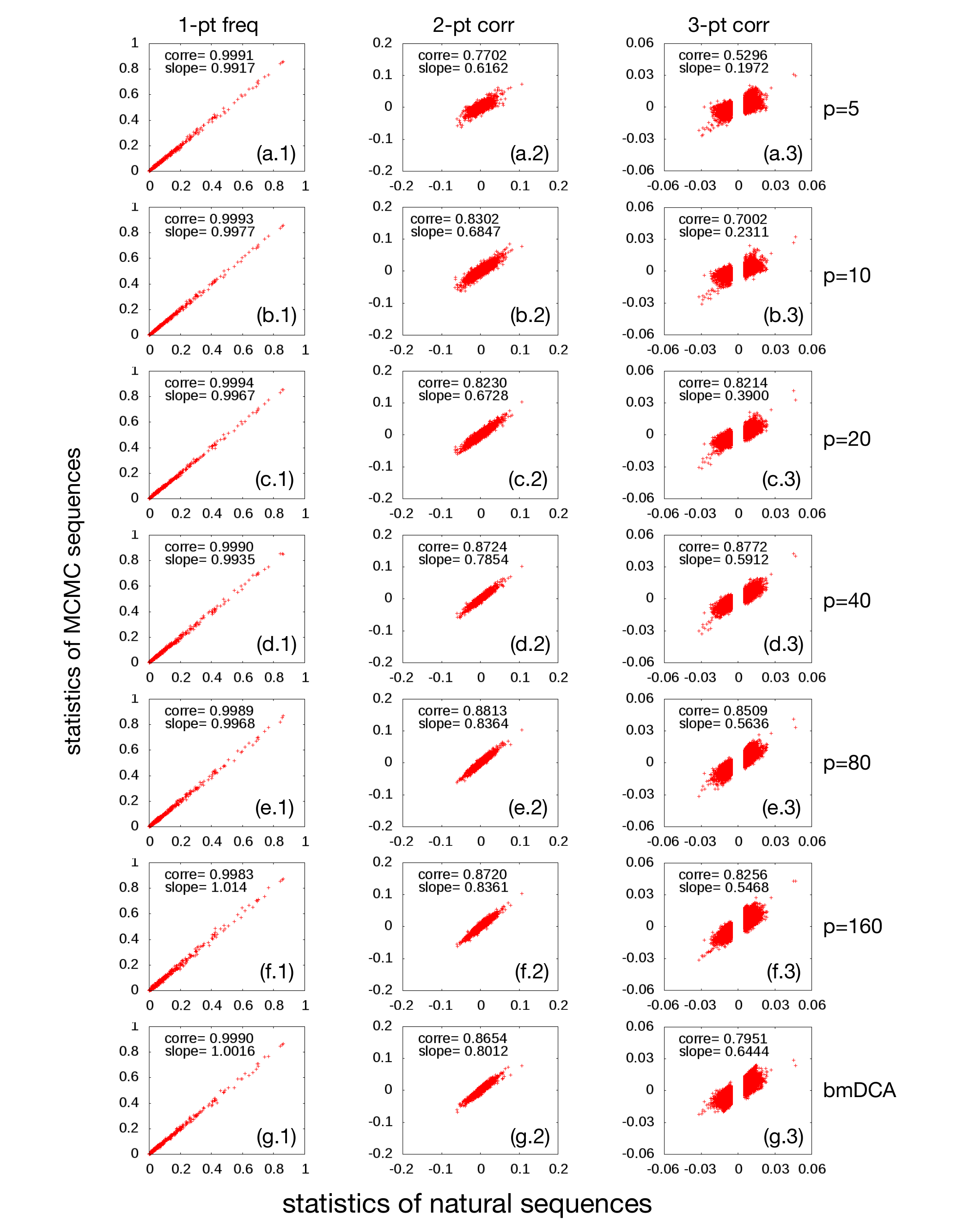}
          \caption{Same as Fig.~\ref{fig:123pointcorrs}, but for the
            protein family PF00076.
          }
         \label{fig:pf76_1}
\end{figure}

\begin{figure}[htb!]
          \centering
          \includegraphics[width=0.8\textwidth]{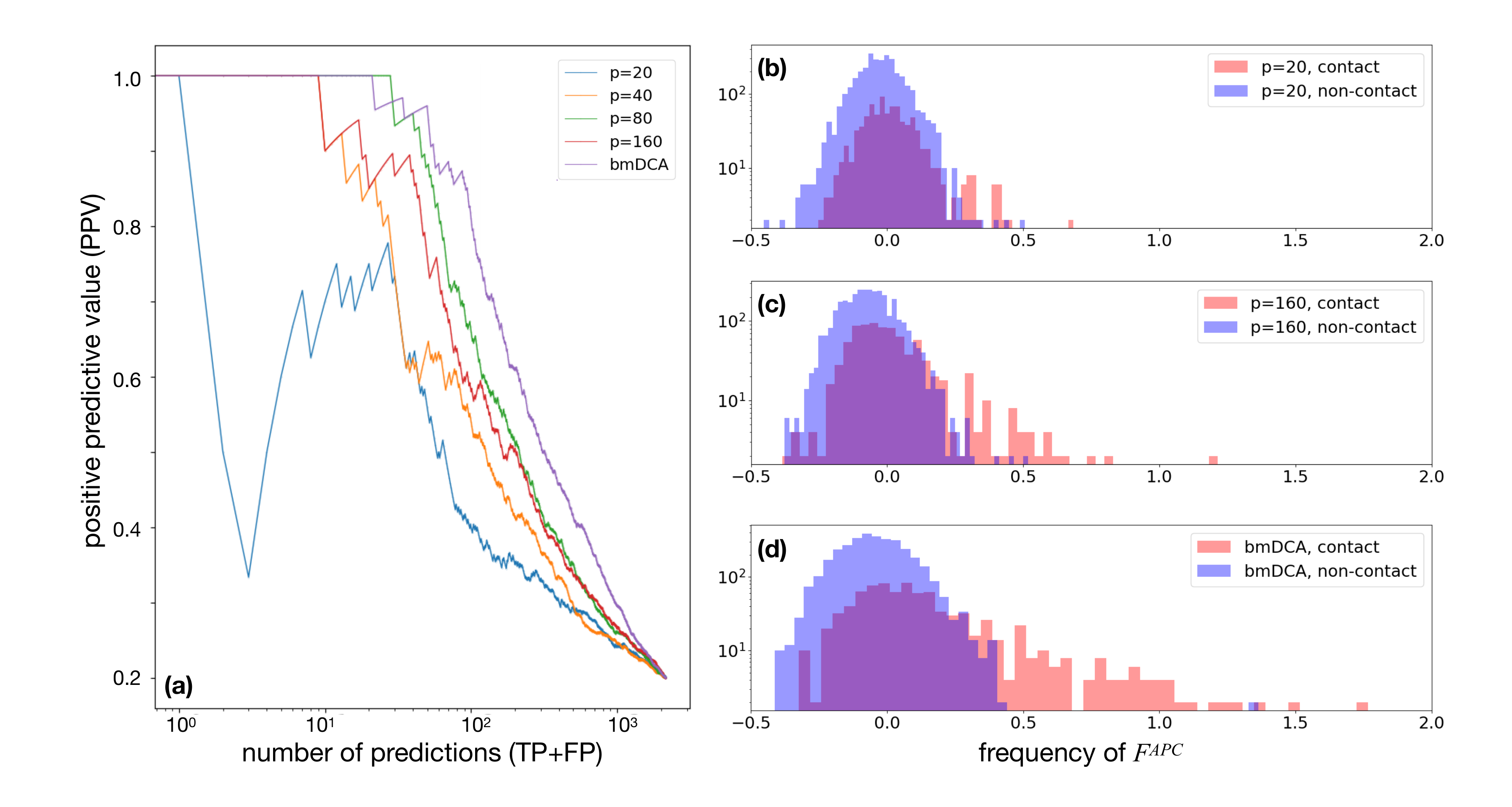}
          \caption{Same as Fig.~\ref{fig:ppv}, but for the
            protein family PF00076.
          }
         \label{fig:pf76_2}
\end{figure}

\begin{figure}[htb!]
          \centering
          \includegraphics[width=0.8\textwidth]{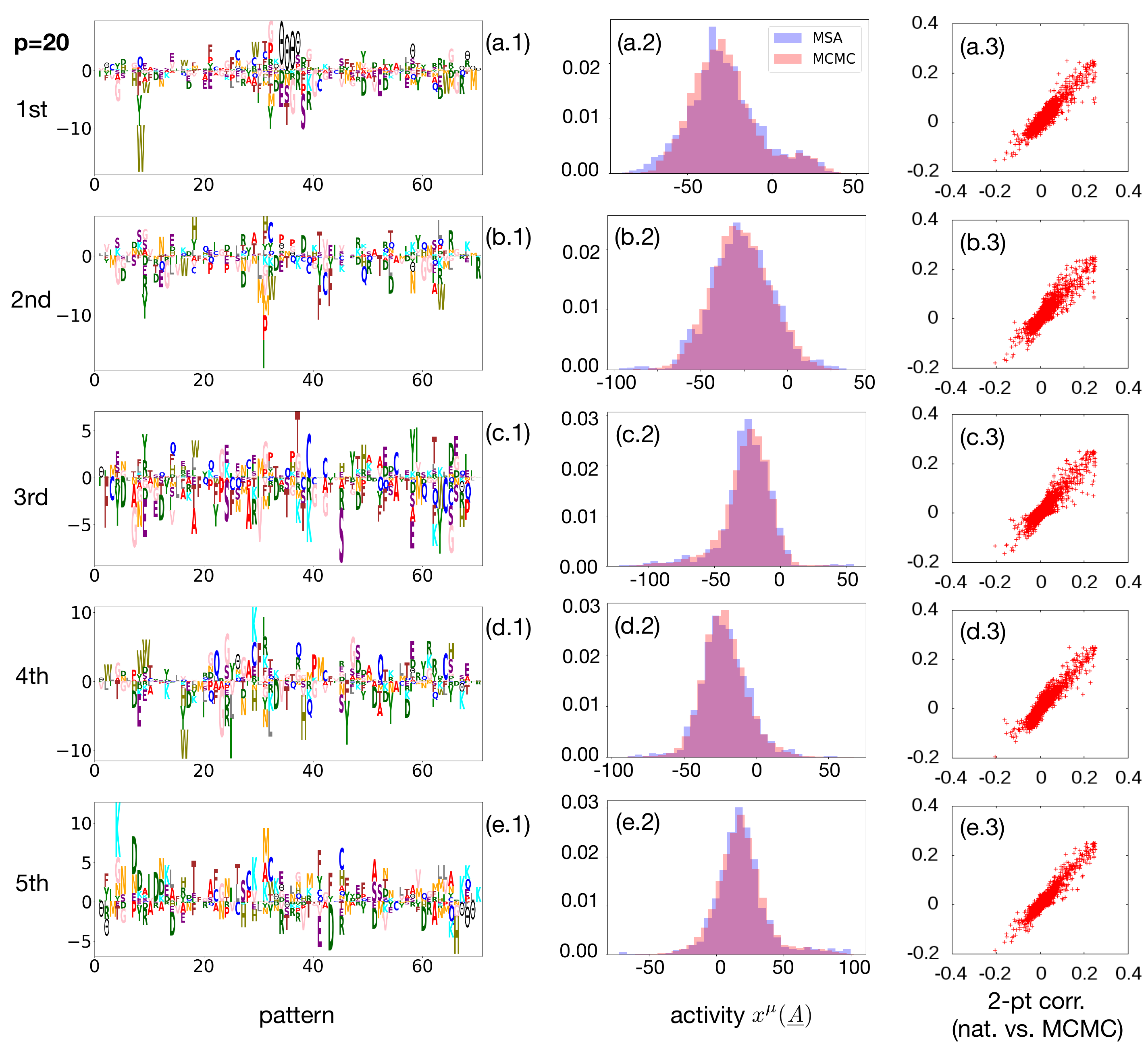}
          \caption{Same as Fig.~\ref{fig:first_patterns}, but for the
            protein family PF00076.
          }
         \label{fig:pf76_3}
\end{figure}

Figs.~\ref{fig:pf76_1}, \ref{fig:pf76_2} et \ref{fig:pf76_3} display
the major results for the PF00076 protein family. PPV curves are
calculated using PDB ID 2x1a \cite{pancevac2010structure}.

\section{Notes and details on inference methods}
\label{app:cd}

\subsection{Regularization}

In case of limited data but many parameters, i.e.~the case (Hopfield-)Potts
models for protein families are in, the direct likelihood maximisation
in Eq.~(\ref{eq:ml}) can lead to overfitting effects, causing problems
in sampling and parameter interpretation. To give a simple example, a
rare and therefore unobserved event would be assigned zero
probability, corresponding to (negative) infinite parameter values.

To cope with this problem, regularization is used. Regularization in general
penalizes large (resp. non-zero) parameter values, and can be
justified in  Bayesian inference as a prior distribution acting on the
parameter values. In this paper and following
\cite{tubiana2019learning}, we use a block regularization of the 
form
\begin{equation}
  R(\xi, h) = \eta_0 \sum_{\mu=1}^p \left(
   \sum_{i,a} | \xi_i^\mu(a) |
    \right)^2 + q \eta_0 \sum_{i,a} h_i(a)^2\ ,
\end{equation}  
with $\eta_0$ being a hyperparameter determining
the strength of regularization. This
regularization weakly favors sparsity of the patterns.

We use $\eta_0=\alpha_0L/qM$ with $\alpha_0=0.0525$ as default values
throughout this paper. In the last section of this  appendix, we show
that Hopfield-Potts inference is robust with respect to this choice.

\subsection{Contrastive divergence vs. persistent contrastive
  divergence}

\subsubsection{Contrastive divergence does not reproduce the two-point
statistics}

Contrastive divergence (CD) is a method for training restricted Boltzmann
machines similar to persistent contrastive divergence. Initialized in
the original data, i.e.~the MSA of natural amino-acid sequences, a few
sampling steps are performed in analogy  to Fig.~\ref{fig:rbm_pcd},
the $k$th step is used in the parameter update to approach a solution
of Eq.~(\ref{eq:stationarity}). However, rather than continuing the MCMC
sampling from this sample, the sample is re-initialized in the
original data after each epoch. This has, a priori, advantages and
disadvantages: The sample remains close to a good sample of the model
in CD, but far from a sample of the intermediate model with not yet
converged parameters.

\begin{figure}[htb!]
          \centering
          \includegraphics[width=0.5\textwidth]{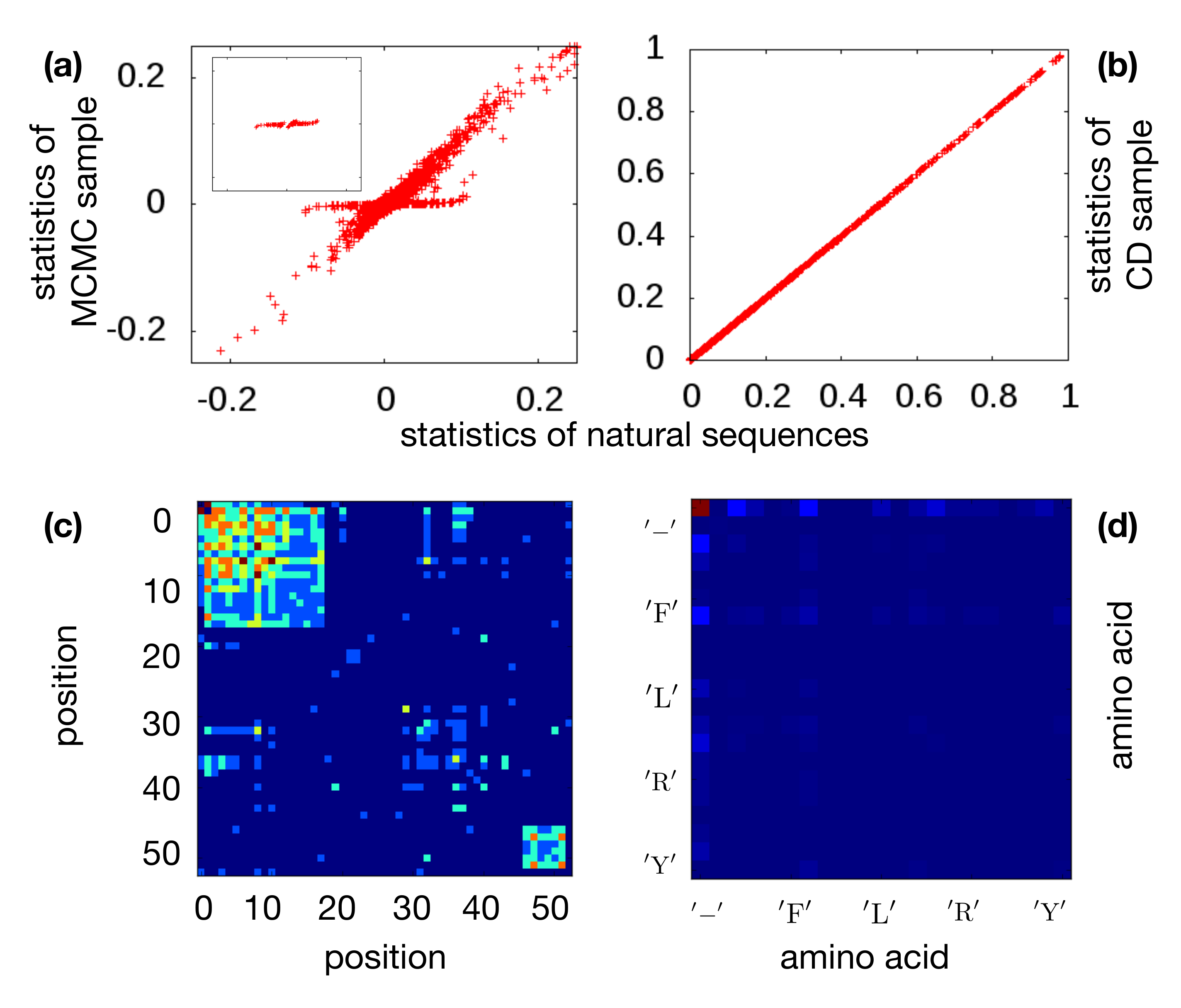}
          \caption{The upper two panels show the statistics (two-point
            connected correlations) of the training data (PF00014) against an {\em i.i.d.}
            MCMC sample extracted from the inferred model, and the CD
            sample used for inference. The perfect coincidence of the
            two in the CD case demonstrates that the CD algorithm is
            converged, contrast is lost despite the fact, that the
            correlations extracted from an {\em i.i.d.} sample do not match
            the empirical ones. To understand this, we have selected
            those $(i,j,a,b)$ with substantial deviations, cf. the
            insert in the first panel, and analyzed their location in
            the protein (first panel) and their amino-acid composition
            (second panel, amino-acids in alphabetical order of one
            letter code $[-,A,C,...Y]$), densities are represented via
            heatmap plots. Location at the extremities od the
            sequences and in gap-gap correlations emerge clearly.
          }
         \label{fig:app_cd}
\end{figure}

As can be seen in Fig.~\ref{fig:app_cd}, after a sufficient number of
epochs the statistics of the CD sample and the training data are
perfectly coherent, the model appears to be converged. However,
the connected two-point correlations are not well reproduced when
resampling the inferred model with standard MCMC. Part of the
empirically non-zero correlations are not reproduced and mistakenly
assigned very small values in the inferred model. 

To understand this observation, we have selected the elements of the
second panel, which show discrepancies between empirical and model
statistics, cf.~the insert in the figure. The corresponding values of
$(i,j,a,b)$ are strongly localized in the beginning and the end of the
protein chain, and correspond to the gap-gap statistics
$c_{ij}(-,-)$. This gives a 
strong hint towards the origin of the problems in CD-based model
inference: gap stretches, which exist in MSA of natural sequences in
particular at the beginning and the end of proteins, due to the local
nature of the alignment algorithm used in Pfam. Those located at the
beginning of the sequence start in position 1, and continue with only
gaps until they are terminated by an amino-acid symbol. They never
start later than in position 1 or include individual internal
amino-acid symbols (analogous for the gap stretches at the end, which
go up to the last position $i=L$).

In CD only a few sampling steps are performed, so stretched gaps in
the initialization tend to be preserved even if the associated gap-gap
couplings are very weak. Basically to remove a gap stretch, an
internal position can not be switched to an amino-acid, but the gap
has to be removed iteratively from one of ts endpoints, namely the one 
inside the sequence (i.e.~not positions 1 or $L$). So, in CD, even
small couplings are thus sufficient to reproduce the gap-gap
statistics. 

If resampling the same model with MCMC, parameters have to be such
that gap-stretches emerge spontaneously during sampling. This requires
quite large couplings, actually in bmDCA gap-gap couplings between
neighboring sites are the largest couplings of the entire Potts
model. Using now the small couplings inferred by CD, these
gap-stretches do not emerge at sufficient frequency, and
correspondingly the positions at the extremities of the sampled
sequences appear less correlated. 

\subsubsection{Persistent contrastive divergence and transient
  oscillations in the two-point statistics}

Persistent contrastive divergence overcomes this sampling issue by not
reinitialising the sample after each epoch, but by continuing the MCMC
exploration in the next epoch, with updated parameters.

\begin{figure}[htb!]
          \centering
          \includegraphics[width=0.5\textwidth]{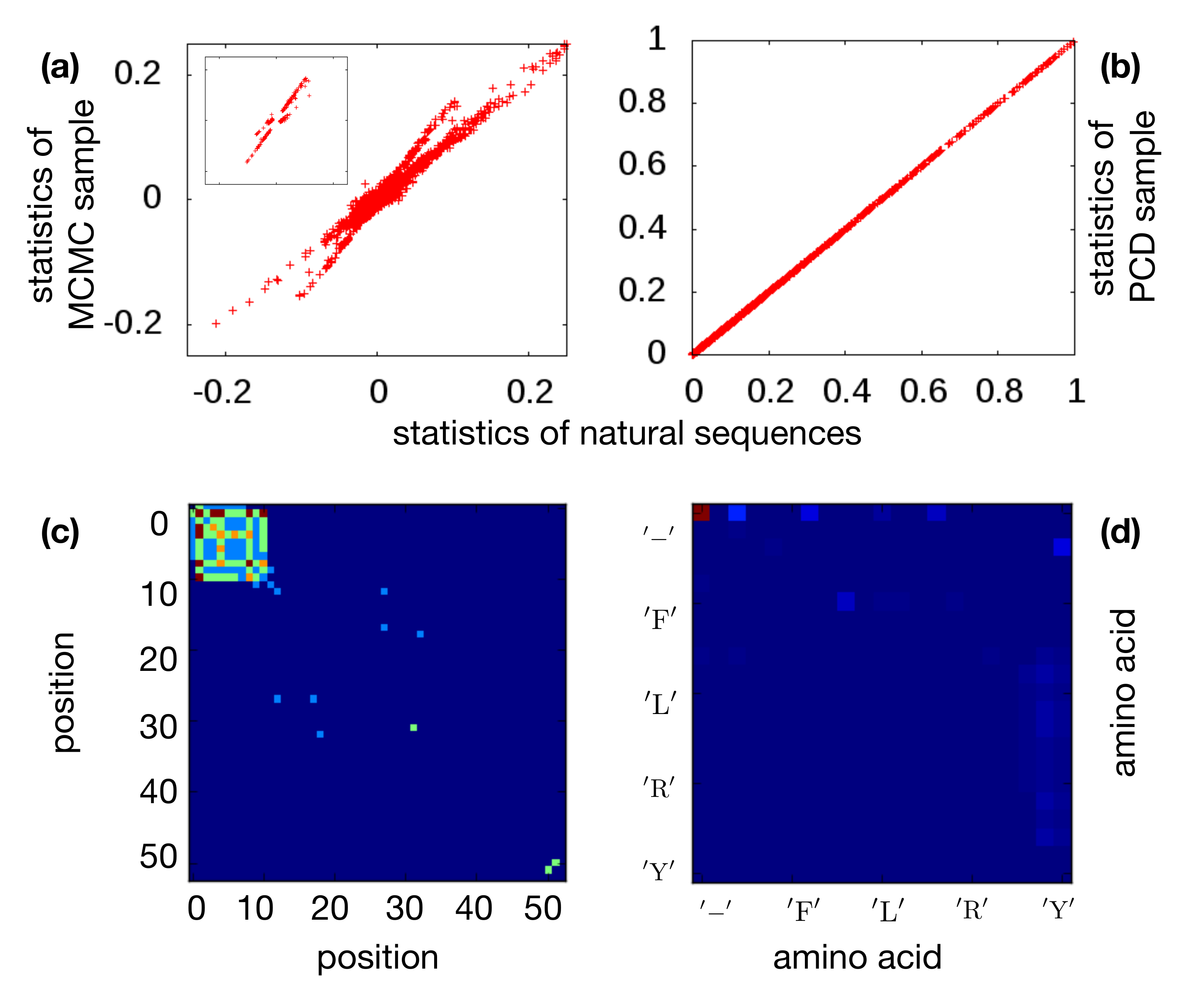}
          \caption{The upper two panels show the statistics (two-point
            connected correlations) of the training data (PF00014) against an {\em i.i.d.}
            MCMC sample extracted from the inferred model, and the PCD
            sample used for inference. As in the CD case, the MCMC
            sample shows larger deviations from the empirical
            observations than the PCD sample, but correlations appear
            overestimated, and contrast in the PCD plot is sufficient
            to drive further evolution of parameters. To understand
            these observation, we have selected
            those $(i,j,a,b)$ with substantial deviations, cf. the
            insert in the first panel, and analyzed their location in
            the protein (first panel) and their amino-acid composition
            (second panel, amino-acids in alphabetical order of one
            letter code $[-,A,C,...Y]$), densities are represented via
            heatmap plots. Location at the extremities od the
            sequences and in gap-gap correlations emerge clearly.
          }
         \label{fig:app_pcd}
\end{figure}

As shown in the main text, PCD can actually be used to infer
parameters, which lead to accurately reproduced two-point
correlations when {\em i.i.d.} samples are generated from the inferred
model. However, during inference we have observed transient
oscillations, cf.~Fig.~\ref{fig:app_pcd} for an epoch where
correlations in a subset of positions and amino acids are
overestimated. An analysis of the of the positions and the spins
involved in these deviations shows, that again gap stretches are
responsible. 

The reason can be understood easily. Initially PCD is not very
different from CD. Gap stretches are present due to the correlation
with the training sample, and only small gap-gap interactions are
learned. However, after a some epochs the sample will loose the
correlation with the training sample. Due to the currently small
gap-gap correlations, gap stretches are lost in the PCD
sample. According to our update rules, the corresponding gap-gap   
couplings will fastly increase. However, due to the few sampling steps performed
in each PCD epoch, this growth will go on even when parameters would
be large enough to generate gap stretches in an {\em i.i.d.} sample. Also
in the PCD sample, gap stretches will now emerge, but due to the
overestimation of parameters, they will be more frequent than in the
training sample, i.e.~parameters start to decrease again. An
oscillation of gap-gap couplings is induced.

The strength of these oscillations can be strongly reduced by removing
samples with large gap stretches from the training data, and train
only on data with limited gaps. If the initial training set was large
enough, the resulting models are even expected to be more precise,
since gap stretches do contain no or little information about the
amino-acid sequences under study. However, if samples are too small,
the suggested pruning procedure may reduce the sample to an
insufficient size for accurate inference. Care has thus to be taken
when removing sequences.

\subsection{Robustness of the results}

\begin{figure}[htb!]
          \centering
          \includegraphics[width=0.7\textwidth]{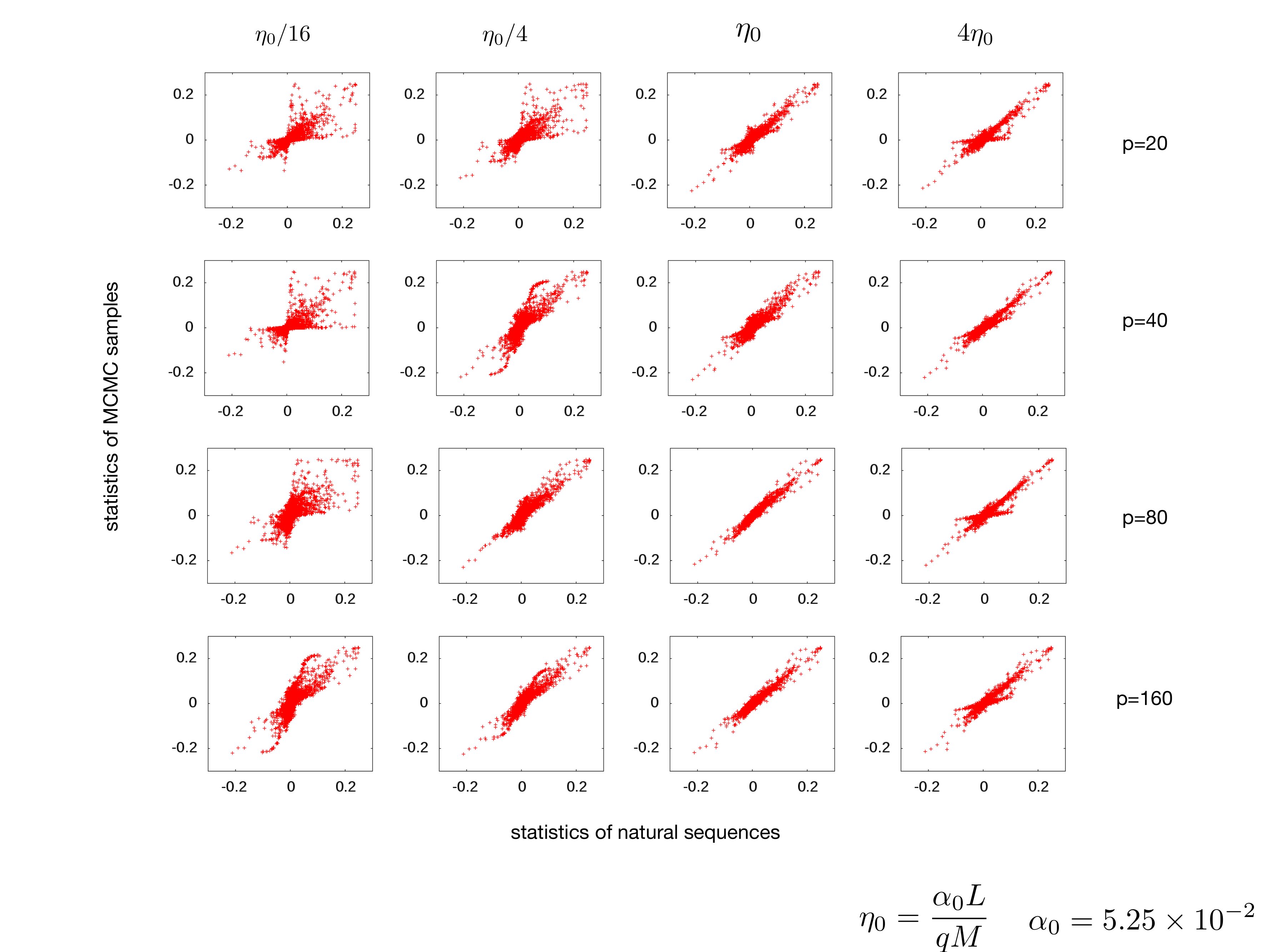}
          \caption{Regularization dependence for CD inference,
            empirical two-point connected correlations (PF00014) are plotted
            against those estimated from the model using an
            {\em i.i.d.} MCMC sample. The regularization strength is varied
            over almost two orders of magnitude, with
            $\eta_0=\alpha_0L/qM$ and $\alpha_0=0.0525$, going from a
            zone of overfitting to one of over-regularization. Results are
            shown for various values of $p$, illustrating a strong
            $p$-dependence of the optimal coupling strength. 
          }
         \label{fig:app_reg_cd}
\end{figure}

\begin{figure}[htb!]
          \centering
          \includegraphics[width=0.7\textwidth]{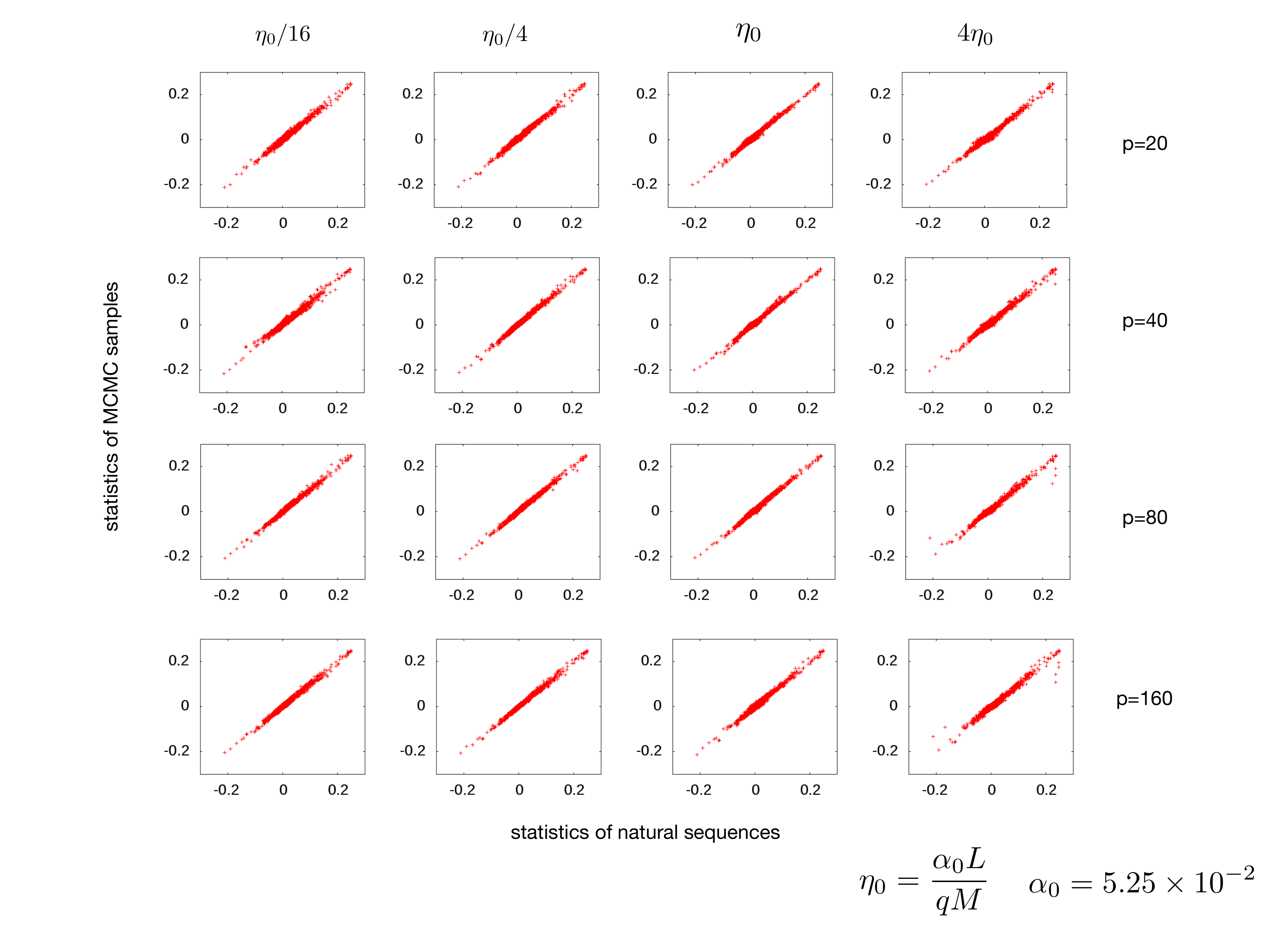}
          \caption{Regularization dependence for PCD inference,
            empirical two-point connected correlations (PF00014) are plotted
            against those estimated from the model using an
            {\em i.i.d.} MCMC sample. The regularization strength is varied
            over almost two orders of magnitude, with
            $\eta_0=\alpha_0L/qM$ and $\alpha_0=0.0525$; results are 
            shown for various values of $p$. We find a strong
            robustness of results with respect to regularization. 
          }
         \label{fig:app_reg_pcd}
\end{figure}

As discussed before, we need to include regularization to avoid
overfitting due to limited data. In Figs.~\ref{fig:app_reg_cd} and
\ref{fig:app_reg_pcd}, we show the dependence of the inference results
due to changes of the regularization strength over roughly two orders
of magnitude.  The first of the two figures shows the
results for CD: empirical connected two-point correlation are compared
with {\em i.i.d.} samples of the corresponding models. We note that the
results depend strongly on the regularization strength. For low
regularization, the correspondence between model and MSA is low, due
to overfitting. At strong regularization, only part of the
correlations is reproduced, we over-regularize and thus underfit the
data. For each protein family, and each number $p$ of patterns, the
regularization strength would have to be tuned. 

For PCD, the situation is fortunately much better, results are found
to be very robust with respect to regularization,
cf.~Fig.~\ref{fig:app_reg_pcd}. This allows us to choose one
regularization strength across protein families and pattern numbers.

\end{document}